\documentclass[onecolumn,draftclsnofoot,12pt]{IEEEtran}
\usepackage{cite}
\usepackage[cmex10]{amsmath}
\usepackage{array}
\usepackage{amsfonts}
\usepackage{mathrsfs}
\usepackage{arydshln}
\usepackage{slashbox}
\usepackage{graphicx}
\usepackage{float}
\usepackage{subfigure}
\usepackage{amssymb}
\usepackage{amsmath}
\usepackage[colorlinks,linkcolor=black,urlcolor=black,anchorcolor=black,citecolor=black,hyperfootnotes=true]{hyperref}
\usepackage{multirow}
\usepackage{multicol}
\usepackage{cite}
\usepackage[subfigure]{graphfig}
\usepackage{xcolor}
\usepackage{setspace}
\newcommand{\mv}[1]{\mbox{\boldmath{$ #1 $}}}
\usepackage[linesnumbered,ruled]{algorithm2e}

\newtheorem{lemma}{\underline{Lemma}}
\newtheorem{proposition}{\underline{Proposition}}

\newcommand{\qed}{\nobreak \ifvmode \relax \else
      \ifdim\lastskip<1.5em \hskip-\lastskip
      \hskip1.5em plus0em minus0.5em \fi \nobreak
      \vrule height0.75em width0.5em depth0.25em\fi}

\begin{document}
\title{{Capacity Characterization for Intelligent Reflecting Surface Aided MIMO Communication}
\author{Shuowen Zhang and Rui Zhang}\thanks{The authors are with the Department of Electrical and Computer Engineering, National University of Singapore (e-mails: \{elezhsh;elezhang\}@nus.edu.sg).}
\author{\IEEEauthorblockN{Shuowen~Zhang, \emph{Member, IEEE} and Rui~Zhang, \emph{Fellow, IEEE}}}}
\maketitle
\begin{abstract}
Intelligent reflecting surface (IRS) is a promising solution to enhance the wireless communication capacity both cost-effectively and energy-efficiently, by properly altering the signal propagation via tuning a large number of passive reflecting units. In this paper, we aim to characterize the fundamental \emph{capacity} limit of IRS-aided point-to-point multiple-input multiple-output (MIMO) communication systems with multi-antenna transmitter and receiver in general, by jointly optimizing the IRS reflection coefficients and the MIMO transmit covariance matrix. First, we consider narrowband transmission under frequency-flat fading channels, and develop an efficient \emph{alternating optimization} algorithm to find a locally optimal solution by iteratively optimizing the transmit covariance matrix or one of the reflection coefficients with the others being fixed. Next, we consider capacity maximization for broadband transmission in a general MIMO orthogonal frequency division multiplexing (OFDM) system under frequency-selective fading channels, where transmit covariance matrices can be optimized for different subcarriers while only one common set of IRS reflection coefficients can be designed to cater to all subcarriers. To tackle this more challenging problem, we propose a new alternating optimization algorithm based on convex relaxation to find a high-quality suboptimal solution. Numerical results show that our proposed algorithms achieve substantially increased capacity compared to traditional MIMO channels without the IRS, and also outperform various benchmark schemes. In particular, it is shown that with the proposed algorithms, various key parameters of the IRS-aided MIMO channel such as channel total power, rank, and condition number can be significantly improved for capacity enhancement.
\end{abstract}
\vspace{-3mm}
\section{Introduction}
\vspace{-1mm}
Driven by the explosive growth of mobile applications such as 8K video streaming and virtual/augmented reality (VR/AR), there is an ever-increasing demand for higher-capacity communications in the fifth-generation (5G) and beyond wireless networks. To achieve this goal, various technologies have been proposed in recent years, among which the most prominent candidates are massive multiple-input multiple-output (MIMO), millimeter wave (mmWave) communication, and ultra-dense networks (UDNs) \cite{5G}. However, although the above technologies are capable of significantly enhancing the wireless network spectral efficiency, they generally require increased energy consumption and hardware cost, due to the need of installing increasingly more active antennas and/or more costly radio frequency (RF) chains operating at higher frequency bands. As such, it still remains an open and challenging problem whether deploying more active components in wireless networks can be a scalable solution for its sustainable capacity growth in the future.

Recently, \emph{intelligent reflecting surface (IRS)} and its various equivalents have emerged as a new and  promising solution to tackle the above challenge \cite{Towards,Survey_Ian,Survey_Renzo,Survey_Basar}. Specifically, IRS is a planar meta-surface equipped with a large number of \emph{passive} reflecting elements connected to a smart controller, which is capable of inducing an independent phase shift and/or amplitude attenuation (collectively termed as ``reflection coefficient'') to the incident signal at each reflecting element in real-time, thereby modifying the wireless channels between one or more pairs of transmitters and receivers to be more favorable for their communications \cite{Towards}. By judiciously designing its reflection coefficients, the signals reflected by IRS can be added either constructively with those via other signal paths to increase the desired signal strength at the receiver, or destructively to mitigate the co-channel interference, thus offering a new degree-of-freedom (DoF) to enhance the communication performance. Since IRSs mainly constitute passive devices without the need of active transmit RF chains \cite{Towards}, they can be densely deployed in wireless networks with low cost and low energy consumption. It is also worth noting that compared to the existing active relays, IRS operates in full-duplex but without requiring additional power for signal amplification/regeneration as well as the sophisticated processing \hbox{for self-interference cancellation \cite{Towards}.}

However, new challenges also arise in the design and implementation of IRS-aided wireless systems. First, to fully exploit the new DoF brought by IRS, the IRS reflection coefficients need to be optimally designed, which requires accurate knowledge of the channel state information (CSI) on the new IRS-related channels with the transmitters and receivers. In practice, this is a difficult task since IRS elements generally cannot transmit/receive signals due to the lack of transmit/receive RF chains. As a result, the conventional pilot-assisted channel estimation is not directly applicable. Moreover, the total number of IRS-related channels increases rapidly with the number of IRS reflecting elements, especially when there are multiple antennas at the transmitter/receiver. To overcome this challenge, \cite{Beyond,CE_Taha} advocated a channel sensing based approach for CSI acquisition by deploying dedicated receive RF chains (or sensors) at IRS, which, however, increases its implementation cost. Alternatively, even without any receive RF chains at IRS, \cite{CE_Johansson,CE_Yuan,Protocol,CE_B} proposed to estimate the transmitter-IRS-receiver concatenated channel based on the training signals sent by the transmitter/receiver as well as the channel reciprocity between the forward and reverse links, for both frequency-flat and frequency-selective fading channels. For example, to reduce the training overhead required for estimating the large number of IRS-involved channel coefficients, a novel grouping-based method was proposed in \cite{Protocol} where only the ``combined channel'' for each IRS element group consisting of multiple adjacent elements needs to be estimated, by exploiting the channel correlations over adjacent elements.

Second, based on the available CSI, how to optimize the \emph{IRS reflection coefficients} (also termed as ``passive beamforming'' design) to maximally reap the IRS performance gains is another crucial problem, which has been studied under various system and channel setups  \cite{Joint_Active,XH_ICCC,TWC_Yuen,MUMISO_Liang,MUMISO_Slim,Discrete_Yuen,Discrete,OFDM,Protocol,CE_B,Statistical_Jin}. 
Furthermore, IRS has been jointly designed with other existing technologies, such as non-orthogonal multiple access (NOMA) \cite{NOMA_Shi,NOMA_Yang,Ding_NOMA}, physical-layer security \cite{Security_GC,Security_XR,Security_Chen,Security_XH,Security_Xu}, and simultaneous wireless information and power transfer (SWIPT) \cite{SWIPT_Wu,Pan_SWIPT}.

It is worth noting that the existing works on IRS-aided communication mainly focused on single-input single-output (SISO) or multiple-input single-output (MISO) systems with single-antenna receivers. However, there has been very limited work on \emph{IRS-aided MIMO communication} with multiple antennas at both the transmitter and the receiver, while only a couple of papers appeared recently \cite{MIMO_Slim,MIMO_Pan}. In particular, the \emph{characterization of the capacity limit} of IRS-aided MIMO communication still remains open, which requires the \emph{joint optimization of IRS reflection coefficients and MIMO transmit covariance matrix}, and thus is more challenging than the traditional MIMO channel capacity characterization \cite{Fundamental} without the IRS reflection. Note that this problem is also more difficult to solve as compared to that in IRS-aided SISO/MISO communications with single data stream transmission only, since the MIMO channel capacity is generally achieved by transmitting \emph{multiple data streams} in parallel (i.e., spatial multiplexing), thus the reflection coefficients need to be properly designed to optimally balance the channel gains for multiple spatial data streams so as to maximize their sum-rate. To the best of our knowledge, this problem has not been fully addressed yet (e.g., in \cite{MIMO_Slim},\cite{MIMO_Pan}), even for the point-to-point IRS-aided MIMO communication, under both frequency-flat and frequency-selective fading channels, which thus motivates this work.

\begin{figure}[t]
	\centering
	\includegraphics[width=9cm]{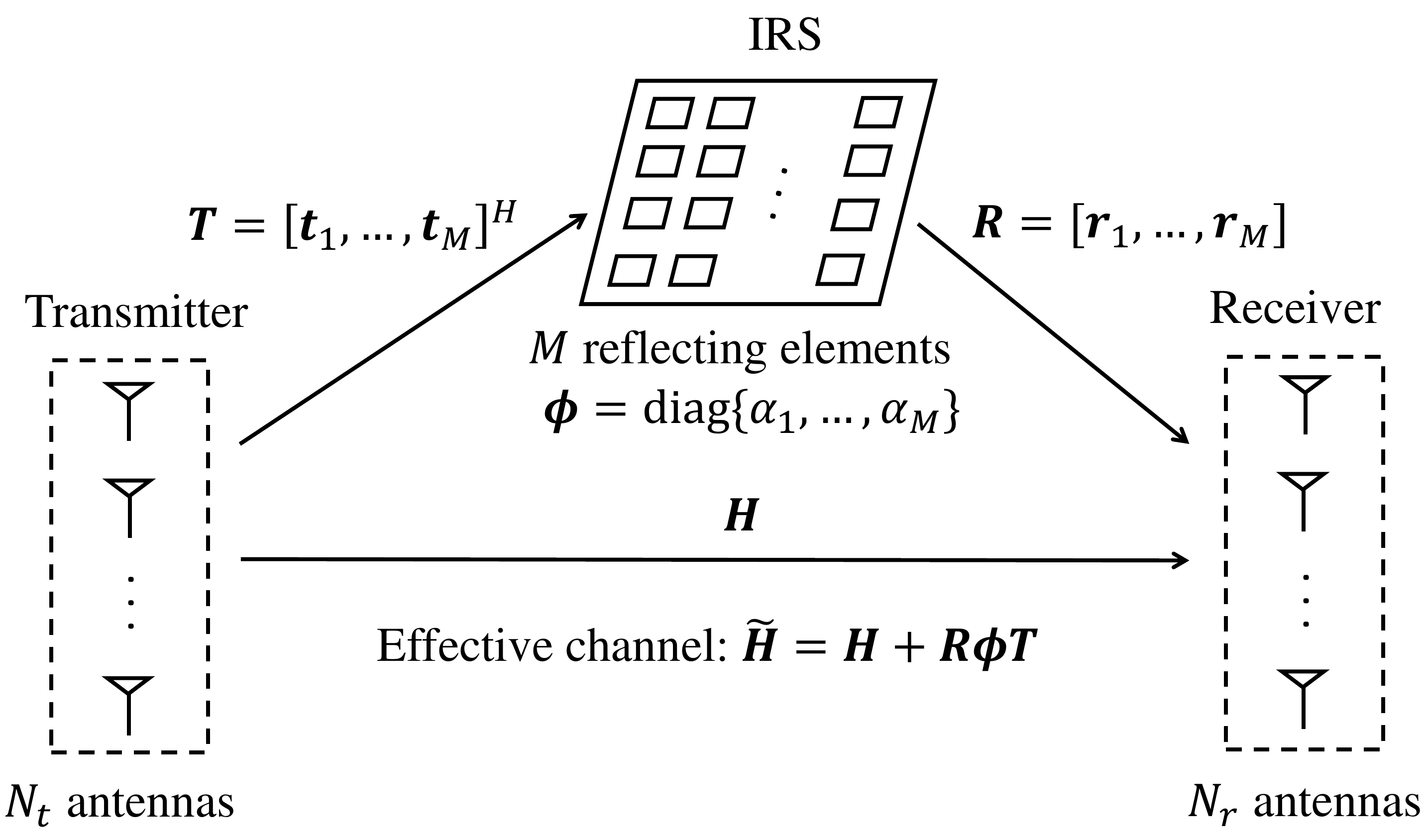}
	\vspace{-3mm}
	\caption{Diagram of an IRS-aided MIMO communication system.}
	\label{IRS}
	\vspace{-9mm}
\end{figure}
In this paper, we study the joint IRS reflection coefficient and transmit covariance matrix optimization for maximizing the capacity of a point-to-point IRS-aided MIMO system with multiple antennas at both the transmitter and the receiver, as illustrated in Fig. \ref{IRS}. To characterize the fundamental capacity limit, we consider that perfect CSI of all channels involved in Fig. \ref{IRS} is available at both the transmitter and the receiver by assuming that the CSI has been accurately acquired via the techniques proposed in e.g., \cite{Beyond,CE_Taha,CE_Johansson,CE_Yuan,Protocol,CE_B}. Moreover, to reduce the implementation complexity of IRS,\footnote{In practice, dynamic change of the resistor load connected to each reflecting element is needed to adjust the reflection amplitude \cite{resistor}, which, however, is difficult to implement in real-time with separate phase-shift control.} we consider that the amplitude of all its reflection coefficients is fixed as the maximum value of one \cite{Towards}. Our main contributions are summarized as follows.
\begin{itemize}
	\item First, we investigate the capacity maximization problem for a narrowband MIMO system under \emph{frequency-flat} channels, which is however non-convex and thus difficult to solve. By exploring the structure of the MIMO capacity expression, we develop an \emph{alternating optimization} algorithm by iteratively optimizing one of the reflection coefficients or the transmit covariance matrix with the others being fixed. We derive the \emph{optimal} solution to each subproblem for optimizing one of these variables in \emph{closed-form}, which greatly reduces the computational complexity. It is shown that the proposed algorithm is guaranteed to converge to at least a \emph{locally optimal} solution.
	\item Moreover, we derive the IRS-aided MIMO channel capacities in the asymptotically low signal-to-noise ratio (SNR) regime and high-SNR regime, respectively, and propose two alternative algorithms for solving the capacity maximization problems in these two cases with lower complexity. In addition, we further simplify the algorithms for the capacity maximization in the special cases of MISO and single-input multiple-output (SIMO) channels.
	\item Next, we consider the general broadband MIMO orthogonal frequency division multiplexing (OFDM) system under \emph{frequency-selective} channels. In this case, individual transmit covariance matrices can be designed for different OFDM subcarriers, while only a common set of IRS reflection coefficients can be designed to cater to all the subcarriers, due to the lack of ``frequency-selective'' passive beamforming capability at the IRS.\footnote{Note that with no receive RF chains and thus no baseband signal processing, IRS can only reflect the broadband signal with ``frequency-flat'' reflection coefficients, which is different from the conventional digital beamforming that can be designed for different frequency sub-bands.} This thus renders the capacity maximization problem more difficult to solve than that in the narrowband MIMO case. By leveraging the convex relaxation technique, we propose a new alternating optimization algorithm for finding a high-quality suboptimal solution in this case.
	\item Finally, we provide extensive numerical results to validate the performance advantages of our proposed alternating optimization algorithms over other benchmark schemes with or without IRS, under both frequency-flat and frequency-selective fading channels. In particular, it is shown that by judiciously designing the IRS reflection coefficients, various key parameters of the IRS-aided MIMO channel such as channel total power, rank and condition number can be significantly improved for capacity enhancement, to draw useful insights into the proposed designs.
\end{itemize}

The rest of this paper is organized as follows. Section \ref{sec_sys} presents the system model and the problem formulation. Section \ref{sec_solution} proposes an alternating optimization algorithm for solving the formulated problem in frequency-flat channels, under different setups. Section \ref{sec_OFDM} extends the proposed algorithm to MIMO-OFDM communication under frequency-selective channels. Numerical results and their pertinent discussions are presented in Section \ref{sec_num}. Finally, Section \ref{sec_conclusion} concludes this paper.

\emph{Notations:} Vectors and matrices are denoted by boldface lower-case letters and boldface upper-case letters, respectively. $|z|$, $z^*$, $\arg\{z\}$, and $\mathfrak{Re}\{z\}$ denote the absolute value, conjugate, angle, and real part of a complex number $z$, respectively. For a complex vector $\mv{x}$, $\|{\mv{x}}\|$ and $x_k$ denote the $l_2$-norm and the $k$th element, respectively, and $\mathrm{diag}\{{\mv{x}}\}$ denotes a square diagonal matrix with the elements of $\mv{x}$ on its main diagonal. $\mathbb{C}^{M\times N}$ denotes the space of $M\times N$ complex matrices, and $\mathbb{R}$ denotes the space of real numbers. $\mv{I}_M$ denotes an $M\times M$ identity matrix, and ${\mv{0}}$ denotes an all-zero matrix with appropriate dimension. For an $M\times N$ matrix $\mv{A}$, $\mv{A}^T$ and $\mv{A}^H$ denote its transpose and conjugate transpose, respectively; $\mathrm{rank}({\mv{A}})$, $[{\mv{A}}]_{i,j}$, and $\|{\mv{A}}\|_F$ denote the rank, $(i,j)$-th element, and Frobenius norm of $\mv{A}$, respectively. For a square matrix $\mv{S}$, $\det(\mv{S})$, $\mathrm{tr}({\mv{S}})$, and $\mv{S}^{-1}$ denote its determinant, trace, and inverse, respectively, and $\mv{S}\succeq {\mv{0}}$ means that $\mv{S}$ is positive semi-definite. The distribution of a circularly symmetric complex Gaussian (CSCG) random variable with mean $\mu$ and variance $\sigma^2$ is denoted by $\mathcal{CN}(\mu,\sigma^2)$; and $\sim$ stands for ``distributed as''. $\mathbb{E}[\cdot]$ denotes the statistical expectation. \hbox{$\mathcal{O}(\cdot)$ denotes the standard big-O notation.}
\vspace{-1mm}
\section{System Model and Problem Formulation}\label{sec_sys}
\vspace{-1mm}
We consider a MIMO communication system with $N_t\geq 1$ antennas at the transmitter and $N_r\geq 1$ antennas at the receiver, as illustrated in Fig. \ref{IRS}, where an IRS equipped with $M$ passive reflecting elements is deployed to enhance the MIMO communication performance. Each element of the IRS is able to re-scatter the signal at the IRS with an individual reflection coefficient, which can be dynamically adjusted by the IRS controller for desired signal reflection. Specifically, let $\alpha_m\in \mathbb{C}$ denote the reflection coefficient of the $m$th IRS element, which is assumed to satisfy $|\alpha_m|=1,\forall m=1,...,M$, while the phase of each $\alpha_m$ can be flexibly adjusted in $[0,2\pi)$ \cite{metamaterial}.\footnote{To characterize the capacity limit of IRS-aided MIMO systems, we assume that the phase-shift by each IRS element can be continuously adjusted, while the results of this paper can be readily extended to the practical setup with discrete phase-shift levels \cite{Discrete_Yuen,Discrete}.}

We assume quasi-static block-fading channels, and focus on one particular fading block where all the channels involved in Fig. \ref{IRS} remain approximately constant. For the purpose of exposition, we will first consider the narrowband transmission over frequency-flat channels in Sections \ref{sec_sys} and \ref{sec_solution}, and then extend the results to the broadband transmission over frequency-selective channels in Section \ref{sec_OFDM}. Denote $\mv{H}\in \mathbb{C}^{N_r\times N_t}$ as the complex baseband channel matrix for the direct link from the transmitter to the receiver, $\mv{T}\in \mathbb{C}^{M\times N_t}$ as that from the transmitter to the IRS, and $\mv{R}\in \mathbb{C}^{N_r\times M}$ as that from the IRS to the receiver. Let $\mv{\phi}\in \mathbb{C}^{M\times M}$ denote the diagonal reflection matrix of the IRS, with $\mv{\phi}=\mathrm{diag}\{\alpha_1,...,\alpha_M\}$. We assume that the signal reflected by the IRS more than once is of negligible power due to the high path loss and thus can be ignored. Therefore, the effective MIMO channel matrix from the transmitter to the receiver is given by $\tilde{\mv{H}}={\mv{H}}+{\mv{R\phi T}}$. 

Let $\mv{x}\in \mathbb{C}^{N_t\times 1}$ denote the transmitted signal vector. The transmit signal covariance matrix is thus defined as $\mv{Q}\overset{\Delta}{=}\mathbb{E}[{\mv{xx}}^H]\in \mathbb{C}^{N_t\times N_t}$, with $\mv{Q}\succeq {\mv{0}}$. We consider an average sum power constraint at the transmitter given by $\mathbb{E}[\|{\mv{x}}\|^2]\leq P$, which is equivalent to $\mathrm{tr}({\mv{Q}})\leq P$. The received signal vector denoted as $\mv{y}\in \mathbb{C}^{N_r\times 1}$ is given by
\vspace{-3mm}\begin{equation}\label{transmission}
{\mv{y}}=\tilde{\mv{H}}{\mv{x}}+\mv{z}=({\mv{H}}+{\mv{R\phi T}}){\mv{x}}+\mv{z},
\vspace{-3mm}\end{equation}
where $\mv{z}\sim \mathcal{CN}(0,\sigma^2\mv{I}_{N_r})$ denotes the independent CSCG noise vector at the receiver, with $\sigma^2$ denoting the average noise power. To reveal the fundamental capacity limit of IRS-aided MIMO communication, we assume that perfect CSI is available at both the transmitter and receiver. The MIMO channel capacity is thus given by
\vspace{-2mm}
\begin{equation}\label{capacity}
C=\underset{\small{\mv{Q}}:\mathrm{tr}(\small{\mv{Q}})\leq P,\small{\mv{Q}}\succeq {\mv{0}}}{\max}\ \log_2\det\left({\mv{I}}_{N_r}+\frac{1}{\sigma^2}\tilde{\mv{H}}\mv{Q}\tilde{\mv{H}}^H\right)
\vspace{-2mm}
\end{equation}
in bits per second per Hertz (bps/Hz). It is worth noting that different from the conventional MIMO channel without the IRS, i.e., $\tilde{\mv{H}}={\mv{H}}$, for which the capacity is solely determined by the channel matrix $\mv{H}$, the capacity for the IRS-aided MIMO channel shown in (\ref{capacity}) is also dependent on the IRS reflection matrix $\mv{\phi}$, since it influences the effective channel matrix $\tilde{\mv{H}}$ as well as the resultant optimal transmit covariance matrix $\mv{Q}$. 

Motivated by the above, we aim to maximize the capacity of an IRS-aided MIMO channel by jointly optimizing the IRS reflection matrix $\mv{\phi}$ and the transmit covariance matrix $\mv{Q}$, subject to uni-modular constraints on the reflection coefficients and a sum power constraint at the transmitter. The optimization problem is formulated as
\vspace{-2mm}
\begin{align}
\mbox{(P1)}\quad \underset{\mv{\phi},{\mv{Q}}}{\max}\  &\log_2\det\left({\mv{I}}_{N_r}+\frac{1}{\sigma^2}\tilde{\mv{H}}\mv{Q}\tilde{\mv{H}}^H\right)\label{P1obj}\\[-2mm]
\mbox{s.t.}\quad & {\mv{\phi}}=\mathrm{diag}\{\alpha_1,...,\alpha_M\}\label{P1c1}\\[-2mm]
&|\alpha_m|= 1,\quad m=1,...,M\label{P1c2}\\[-2mm]
&\mathrm{tr}(\mv{Q})\leq P\\[-2mm]
& {\mv{Q}} \succeq {\mv{0}}.
\end{align}
\vspace{-11mm}

Note that Problem (P1) is a non-convex optimization problem since the objective function can be shown to be non-concave over the reflection matrix $\mv{\phi}$, and the uni-modular constraint on each reflection coefficient $\alpha_m$ in (\ref{P1c2}) is also non-convex. Moreover, the transmit covariance matrix ${\mv{Q}}$ is coupled with $\mv{\phi}$ in the objective function of (P1), which makes (P1) more difficult to solve. It is worth noting that although uni-modular constraints have been considered in the designs of constant envelope precoding and hybrid analog/digital precoding at the transmitter (see, e.g., \cite{CE_MIMO,mmWave_Heath,mmWave_Yu}), the existing designs are not applicable to solving (P1) due to the different rate expressions in terms of the uni-modular variables. In the next section, we solve (P1) by exploiting its unique structure.
\vspace{-2mm}
\section{Proposed Solution to Problem (P1)}\label{sec_solution}
\vspace{-1mm}
In this section, we propose an alternating optimization algorithm for solving (P1). Specifically, we first transform the objective function of (P1) into a more tractable form in terms of the optimization variables in $\{\alpha_m\}_{m=1}^M\cup\{{\mv{Q}}\}$, based on which we then solve two subproblems of (P1), for optimizing respectively the transmit covariance matrix $\mv{Q}$ or one reflection coefficient $\alpha_m$ in $\mv{\phi}$ with all the other variables being fixed. We derive the \emph{optimal} solutions to both subproblems in \emph{closed-form}, which enable an efficient alternating optimization algorithm to obtain a \emph{locally optimal} solution to (P1) by iteratively solving these subproblems. Next, we derive more tractable expressions of the MIMO channel capacity for the asymptotically low-SNR and high-SNR regimes, based on which we propose two alternative low-complexity solutions to (P1), respectively. Finally, we consider the special cases of (P1) with single-antenna transmitter/receiver, and propose further simplified algorithms in these special cases.
\vspace{-1mm}
\subsection{Alternating Optimization}\label{sec_framework}
\vspace{-1mm}
In this subsection, we introduce the framework of our proposed alternating optimization for solving (P1). Our main idea is to iteratively solve a series of subproblems of (P1), each aiming to optimize one single variable in $\{\alpha_m\}_{m=1}^M\cup\{{\mv{Q}}\}$ with the other $M$ variables being fixed. To this end, we first provide a more tractable expression for the objective function of (P1) in (\ref{P1obj}) in terms of $\mv{Q}$ and $\{\alpha_m\}_{m=1}^M$. Note that (\ref{P1obj}) is the logarithm determinant of a linear function of $\mv{Q}$, while its relationship with $\alpha_m$'s is rather implicit. Thus, we propose to rewrite (\ref{P1obj}) as an explicit function over $\alpha_m$'s. Denote $\mv{R}=[{\mv{r}}_1,...,{\mv{r}}_M]$ and $\mv{T}=[{\mv{t}}_1,...,{\mv{t}}_M]^H$, where ${\mv{r}}_m\in \mathbb{C}^{N_r\times 1}$ and ${\mv{t}}_m\in \mathbb{C}^{N_t\times 1}$. Then, the effective MIMO channel can be rewritten as
\vspace{-2mm}
\begin{equation}\label{effchannel}
\tilde{\mv{H}}=\mv{H}+\sum_{m=1}^M \alpha_m{\mv{r}}_m{\mv{t}}_m^H.
\vspace{-2mm}
\end{equation} 
Notice from (\ref{effchannel}) that the effective channel is in fact the summation of the direct channel matrix $\mv{H}$ and $M$ rank-one matrices ${\mv{r}}_m{\mv{t}}_m^H$'s each multiplied by a reflection coefficient $\alpha_m$, which is a unique structure of IRS-aided MIMO channel and implies that $\{\alpha_m\}_{m=1}^M$ should be designed to strike an optimal balance between the $M+1$ matrices for maximizing the channel capacity.

Furthermore, denote ${\mv{Q}}={\mv{U}}_Q{\mv{\Sigma}}_Q{\mv{U}}_Q^H$ as the eigenvalue decomposition (EVD) of $\mv{Q}$, where ${\mv{U}}_Q\in \mathbb{C}^{N_t\times N_t}$ and $\mv{\Sigma}_Q\in \mathbb{C}^{N_t\times N_t}$. Note that since ${\mv{Q}}$ is a positive semi-definite matrix, all the diagonal elements in ${\mv{\Sigma}_Q}$ are non-negative real numbers. Based on this, we define ${\mv{H}}'={\mv{H}}{\mv{U}}_Q{\mv{\Sigma}}_Q^{\frac{1}{2}}\in \mathbb{C}^{N_r\times N_t}$, ${\mv{T}}'={\mv{T}}{\mv{U}}_Q{\mv{\Sigma}}_Q^{\frac{1}{2}}=[{\mv{t}}_1',...,{\mv{t}}_M']^H\in \mathbb{C}^{M\times N_t}$, where ${\mv{t}}_m'={\mv{t}}_m{\mv{U}}_Q{\mv{\Sigma}}_Q^{\frac{1}{2}}\in \mathbb{C}^{N_t\times 1}$. Therefore, the objective function of (P1) can be rewritten as
\begin{align}\label{obj1}
f\overset{\Delta}{=}&\log_2\det\left({\mv{I}}_{N_r}+\frac{1}{\sigma^2}\tilde{\mv{H}}\mv{Q}\tilde{\mv{H}}^H\right)=\log_2\det\left({\mv{I}}_{N_r}+\frac{1}{\sigma^2}\left(\tilde{\mv{H}}{\mv{U}}_Q{\mv{\Sigma}}_Q^{\frac{1}{2}}\right)\left(\tilde{\mv{H}}{\mv{U}}_Q{\mv{\Sigma}}_Q^{\frac{1}{2}}\right)^H\right)\nonumber\\[-2mm]
=&\log_2\det\left({\mv{I}}_{N_r}+\frac{1}{\sigma^2}({\mv{H}}'+{\mv{R\phi T}}')({\mv{H}}'+{\mv{R\phi T}}')^H\right)\nonumber\\[-2mm]
=&\log_2\det\left({\mv{I}}_{N_r}+\frac{1}{\sigma^2}\left({\mv{H}}'+\sum_{i=1}^M\alpha_i{\mv{r}}_i{\mv{t}}_i^{'H}\right)\left({\mv{H}}'+\sum_{i=1}^M\alpha_i{\mv{r}}_i{\mv{t}}_i^{'H}\right)^H\right)\nonumber\\[-2mm]
\overset{(a)}{=}&\log_2\det\Bigg({\mv{I}}_{N_r}+\frac{1}{\sigma^2}\mv{H}'\mv{H}^{'H}+\frac{1}{\sigma^2}\sum_{i=1}^M{\mv{r}}_i{\mv{t}}_i^{'H}{\mv{t}}_i'{\mv{r}}_i^H+\frac{1}{\sigma^2}\sum_{i=1}^M\sum_{j=1,j\neq i}^M\alpha_i\alpha_j^*{\mv{r}}_i{\mv{t}}_i^{'H}{\mv{t}}_j'{\mv{r}}_j^H\nonumber\\[-2mm]
&\qquad\qquad +\frac{1}{\sigma^2}\sum_{i=1}^M\bigg({\mv{H}}' \alpha_i^*{\mv{t}}_i'{\mv{r}}_i^H+ \alpha_i{\mv{r}}_i{\mv{t}}_i^{'H}{\mv{H}}^{'H}\bigg)\Bigg),
\end{align}
where $(a)$ holds due to $|\alpha_m|^2=1,\forall m$. Note that the new objective function of (P1) shown in (\ref{obj1}) is in an explicit form of individual reflection coefficients $\{\alpha_m\}_{m=1}^M$, which facilitates our proposed alternating optimization in the sequel.

With (\ref{obj1}), we are ready to present the two types of subproblems that need to be solved during the alternating optimization, which aim to optimize the transmit covariance matrix $\mv{Q}$ with given $\{\alpha_m\}_{m=1}^M$ or a reflection coefficient $\alpha_m$ with given $\{\alpha_i,i\neq m\}_{i=1}^M\cup\mv{Q}$, elaborated as follows.
\subsubsection{Optimization of $\mv{Q}$ with Given $\{\alpha_m\}_{m=1}^M$}
In this subproblem, we aim to optimize the transmit covariance matrix $\mv{Q}$ with given reflection coefficients $\{\alpha_m\}_{m=1}^M$ or the effective channel $\tilde{\mv{H}}$ in (\ref{effchannel}). Note that with given $\tilde{\mv{H}}$, (P1) is a convex optimization problem over $\mv{Q}$, and the optimal $\mv{Q}$ is given by the \emph{eigenmode transmission} \cite{Fundamental}. Specifically, denote $\tilde{\mv{H}}=\tilde{\mv{U}}\tilde{\mv{\Lambda}}\tilde{\mv{V}}^H$ as the truncated singular value decomposition (SVD) of $\tilde{\mv{H}}$, where $\tilde{\mv{V}}\in \mathbb{C}^{N_t\times D}$, with $D=\mathrm{rank}(\tilde{\mv{H}})\leq \min(N_t,N_r)$ denoting the maximum number of data streams that can be transmitted over $\tilde{\mv{H}}$. The optimal $\mv{Q}$ is thus given by 
\vspace{-3mm}\begin{equation}\label{Qstar}
{\mv{Q}}^\star =\tilde{\mv{V}}\mathrm{diag}\{p_1^\star,...,p_D^\star\}\tilde{\mv{V}}^H,
\vspace{-3mm}\end{equation} 
where $p_i^\star$ denotes the optimal amount of power allocated to the $i$th data stream following the water-filling strategy: $p_i^\star=\max(1/p_0-\sigma^2/[\tilde{\mv{\Lambda}}]_{i,i}^2,0),\ i=1,...,D$, with $p_0$ satisfying $\sum_{i=1}^Dp_i^\star=P$. Hence, the channel capacity with given $\{\alpha_m\}_{m=1}^M$ is $C=\sum_{i=1}^D\log_2\left(1+[\tilde{\mv{\Lambda}}]_{i,i}^2p_i^\star/\sigma^2\right)$.
\subsubsection{Optimization of $\alpha_m$ with Given $\mv{Q}$ and $\{\alpha_i,i\neq m\}_{i=1}^M$}
In this subproblem, we aim to obtain the optimal $\alpha_m$ in (P1) with given $\mv{Q}$ and $\{\alpha_i,i\neq m\}_{i=1}^M$, $\forall m\in \mathcal{M}$, where $\mathcal{M}=\{1,...,M\}$. For ease of exposition, we rewrite the objective function of (P1) in (\ref{obj1}) in the following form with respect to each $\alpha_m$:
\vspace{-3mm}\begin{equation}
f_m\overset{\Delta}{=}\log_2\mathrm{det}\left({\mv{A}}_m+\alpha_m{\mv{B}}_m+\alpha_m^*{\mv{B}}_m^H\right)=f,\quad \forall m\in \mathcal{M},
\vspace{-3mm}\end{equation}
where
\vspace{-2mm}\begin{align}\label{A_mB_m}
{\mv{A}}_m=&{\mv{I}}_{N_r}+\frac{1}{\sigma^2}\left({\mv{H}}'+\sum_{i=1,i\neq m}^M \alpha_i{\mv{r}}_i{\mv{t}}_i^{'H}\right)\left({\mv{H}}'+\sum_{i=1,i\neq m}^M \alpha_i{\mv{r}}_i{\mv{t}}_i^{'H}\right)^H+\frac{1}{\sigma^2}{\mv{r}}_m{\mv{t}}_m^{'H}{\mv{t}}_m'{\mv{r}}_m^H,\forall m\in \mathcal{M}\nonumber\\[-2mm]
{\mv{B}}_m=&\frac{1}{\sigma^2} {\mv{r}}_m{\mv{t}}_m^{'H}\left({\mv{H}}^{'H}+\sum_{i=1,i\neq m}^M {\mv{t}}_i'{\mv{r}}_i^H\alpha_i^*\right),\quad \forall m\in \mathcal{M}.
\end{align}
Therefore, this subproblem can be expressed as
\vspace{-2mm}\begin{align}
\mbox{(P1-m)}\quad \underset{\alpha_m}{\max}\  &\log_2\mathrm{det}({\mv{A}}_m+\alpha_m{\mv{B}}_m+\alpha_m^*{\mv{B}}_m^H)\\[-2mm]
\mbox{s.t.}\quad & |\alpha_m|=1.\label{P1mc1}
\end{align}
\vspace{-11mm}

Notice that ${\mv{A}}_m$ and ${\mv{B}}_m$ are both independent of $\alpha_m$. Hence, the objective function of (P1-m) can be shown to be a concave function over $\alpha_m$. Nevertheless, the uni-modular constraint in (\ref{P1mc1}) is non-convex, which makes (P1-m) still non-convex. In the following, by exploiting the structure of (P1-m), we derive its {\emph{optimal}} solution in {\emph{closed-form}}.
\vspace{-4mm}
\subsection{Optimal Solution to Problem (P1-m)}
\vspace{-2mm}
First, we exploit the structures of $\mv{A}_m$ and $\mv{B}_m$ in the following lemma.
\begin{lemma}\label{lemma_rank}
For any $m\in \mathcal{M}$, $\mathrm{rank}(\mv{A}_m)=N_r$, $\mathrm{rank}(\mv{B}_m)\leq 1$.
\end{lemma}
\begin{IEEEproof}
Note from (\ref{A_mB_m}) that $\mv{A}_m$ is the summation of an identity matrix and two positive semi-definite matrices. Thus, $\mv{A}_m$ is a positive definite matrix with full rank. On the other hand, based on the definition of $\mv{B}_m$ in (\ref{A_mB_m}), we have $\mathrm{rank}(\mv{B}_m)\leq \mathrm{rank}({\mv{r}}_m{\mv{t}}_m^{'H})=1$ \cite{Matrix_Analysis}. This thus completes the proof of Lemma \ref{lemma_rank}.
\end{IEEEproof}

Next, by noting from Lemma \ref{lemma_rank} that ${\mv{A}}_m$ is of full rank and thus invertible, we rewrite the objective function of (P1-m) as
\vspace{-3mm}\begin{equation}\label{f_m1}
f_m=\log_2\det ({\mv{I}}_{N_r}+\alpha_m{\mv{A}}_m^{-1}{\mv{B}}_m+\alpha_m^*{\mv{A}}_m^{-1}{\mv{B}}_m^H)+ \log_2\det ({\mv{A}}_m)\overset{\Delta}{=}f_m'+\log_2\det ({\mv{A}}_m).
\vspace{-3mm}\end{equation}
Based on (\ref{f_m1}), (P1-m) is equivalent to the maximization of $f_m'\overset{\Delta}{=}\log_2\det ({\mv{I}}_{N_r}+\alpha_m{\mv{A}}_m^{-1}{\mv{B}}_m+\alpha_m^*{\mv{A}}_m^{-1}{\mv{B}}_m^H)$ under the constraint in (\ref{P1mc1}) by optimizing $\alpha_m$, which is addressed next. 

Notice that ${\mv{A}}_m^{-1}{\mv{B}}_m$ plays a key role in our new objective function $f_m'$, whose structure is exploited as follows. Specifically, since $\mathrm{rank}(\mv{B}_m)\leq 1$, we have $\mathrm{rank}({\mv{A}}_m^{-1}{\mv{B}}_m)\leq \mathrm{rank}({\mv{B}}_m)\leq 1$. Note that for the case with $\mathrm{rank}({\mv{A}}_m^{-1}\mv{B}_m)=0$, namely, ${\mv{A}}_m^{-1}{\mv{B}}_m={\mv{0}}$, any $\alpha_m$ with $|\alpha_m|=1$ is an optimal solution to (P1-m), whose corresponding optimal value is thus $\log_2\det ({\mv{A}}_m)$. As such, we focus on the case with $\mathrm{rank}({\mv{A}}_m^{-1}\mv{B}_m)=1$ in the next. In this case, ${\mv{A}}_m^{-1}{\mv{B}}_m$ may be either \emph{diagonalizable} or \emph{non-diagonalizable}, which can be determined by the following lemma.
\begin{lemma}\label{lemma_diagonalize}
${\mv{A}}_m^{-1}{\mv{B}}_m$ is diagonalizable if and only if $\mathrm{tr}({\mv{A}}_m^{-1}{\mv{B}}_m)\neq 0$.
\end{lemma}
\begin{IEEEproof}
First, since ${\mv{A}}_m^{-1}{\mv{B}}_m$ is of rank one, we can express it as the multiplication of two vectors as ${\mv{A}}_m^{-1}{\mv{B}}_m={\mv{u}}_m{\mv{v}}_m^H$, where ${\mv{u}}_m\in \mathbb{C}^{N_r\times 1}$ and ${\mv{v}}_m\in \mathbb{C}^{N_r\times 1}$. Then, it follows that ${\mv{A}}_m^{-1}{\mv{B}}_m$ is non-diagonalizable if and only if ${\mv{v}}_m^H{\mv{u}}_m=\mathrm{tr}({\mv{A}}_m^{-1}{\mv{B}}_m)=0$, where it becomes a \textit{nilpotent} matrix \cite{Matrix_Analysis}. This completes the proof of Lemma \ref{lemma_diagonalize}.
\end{IEEEproof}

In the following, we investigate the two cases where ${\mv{A}}_m^{-1}{\mv{B}}_m$ is diagonalizable or non-diagonalizable, and derive the optimal solution for each case, respectively.
\subsubsection{Case I: Diagonalizable ${\mv{A}}_m^{-1}{\mv{B}}_m$}
First, we consider the case where ${\mv{A}}_m^{-1}{\mv{B}}_m$ is diagonalizable, namely, its EVD exists. Since ${\mv{A}}_m^{-1}{\mv{B}}_m$ has rank one, its EVD can be expressed as ${\mv{A}}_m^{-1}{\mv{B}}_m={\mv{U}}_m\mv{\Sigma}_m{\mv{U}}_m^{-1}$, where ${\mv{U}}_m\in \mathbb{C}^{N_r\times N_r}$, and $\mv{\Sigma}_m=\mathrm{diag}\{\lambda_m,0,...,0\}\in \mathbb{C}^{N_r\times N_r}$, with $\lambda_m\in \mathbb{C}$ denoting the sole non-zero eigenvalue of ${\mv{A}}_m^{-1}{\mv{B}}_m$. Therefore, $f_m'$ can be expressed as
\vspace{-2mm}\begin{align}\label{fm'1}
f_m'=&\log_2\det({\mv{I}}_{N_r}+\alpha_m{\mv{U}}_m\mv{\Sigma}_m{\mv{U}}_m^{-1}+\alpha_m^*{\mv{A}}_m^{-1}{\mv{U}}_m^{-1 H}\mv{\Sigma}_m^H{\mv{U}}_m^H{\mv{A}}_m)\nonumber\\[-2mm]
\overset{(b_1)}{=}&\log_2(\det({\mv{U}}_m^{-1})\det({\mv{I}}_{N_r}+\alpha_m{\mv{U}}_m\mv{\Sigma}_m{\mv{U}}_m^{-1}+\alpha_m^*{\mv{A}}_m^{-1}{\mv{U}}_m^{-1 H}\mv{\Sigma}_m^H{\mv{U}}_m^H{\mv{A}}_m)\det({\mv{U}}_m))\nonumber\\[-2mm]
\overset{(b_2)}{=}&\log_2\det({\mv{I}}_{N_r}+\alpha_m\mv{\Sigma}_m+\alpha_m^*{\mv{U}}_m^{-1}{\mv{A}}_m^{-1}{\mv{U}}_m^{-1 H}{\mv{\Sigma}}_m^H{\mv{U}}_m^H{\mv{A}}_m{\mv{U}}_m)\nonumber\\[-2mm]
=&\log_2\det({\mv{I}}_{N_r}+\alpha_m\mv{\Sigma}_m+\alpha_m^*{\mv{V}}_m^{-1}{\mv{\Sigma}}_m^H{\mv{V}}_m),
\end{align}
where $(b_1)$ holds due to $\det({\mv{A}})\det({\mv{A}}^{-1})=1$ for any invertible matrix $\mv{A}$; $(b_2)$ holds due to $\det({\mv{AB}})=\det({\mv{A}})\det({\mv{B}})$ for two equal-sized square matrices $\mv{A}$ and $\mv{B}$; and ${\mv{V}}_m\overset{\Delta}{=}{\mv{U}}_m^H{\mv{A}}_m{\mv{U}}_m$ is a Hermitian matrix with $\mv{V}_m={\mv{V}}_m^H$, since ${\mv{A}}_m$ is a Hermitian matrix according to (\ref{A_mB_m}). Let ${\mv{\nu}}_m\in \mathbb{C}^{N_r\times 1}$ denote the first column of ${\mv{V}}_m^{-1}$ and ${\mv{\nu}}_m^{'T}\in \mathbb{C}^{1\times N_r}$ denote the first row of ${\mv{V}}_m$. Note that it follows that ${\mv{\nu}}_m^{' T}{\mv{\nu}}_m=1$; moreover, let $\nu_{m1}$ and $\nu_{m1}'$ denote the first element in ${\mv{\nu}}_m$ and ${\mv{\nu}}_m^{' T}$, respectively, we have 
$\nu_{m1}\in \mathbb{R}$ and $\nu_{m1}'\in \mathbb{R}$ since both $\mv{V}_m$ and $\mv{V}_m^{-1}$ are Hermitian matrices. Hence, (\ref{fm'1}) can be further simplified as
\vspace{-2mm}
\begin{align}\label{f_m'2}
f_m'=&\log_2\det({\mv{I}}_{N_r}+\alpha_m\mv{\Sigma}_m+\alpha_m^*{\mv{\nu}}_m\lambda_m^*{\mv{\nu}}_m^{' T})\nonumber\\[-2mm]
\overset{(c_1)}{=}&\log_2\det (1+\alpha_m^*\lambda_m^*{\mv{\nu}}_m^{' T}({\mv{I}}_{N_r}+\alpha_m\mv{\Sigma}_m)^{-1}{\mv{\nu}}_m)+\log_2 \det ({\mv{I}}_{N_r}+\alpha_m\mv{\Sigma}_m)\nonumber\\[-2mm]
=& \log_2 \left(\left(1+\alpha_m^*\lambda_m^*{\mv{\nu}}_m^{' T}\left({\mv{I}}_{N_r}-\mathrm{diag}\left\{\frac{\alpha_m\lambda_m}{1+\alpha_m\lambda_m},0,...,0\right\}\right){\mv{\nu}}_m\right)(1+\alpha_m\lambda_m)\right)\nonumber\\[-2mm]
=& \log_2 \left(\left(1+\alpha_m^*\lambda_m^*-\frac{\alpha_m^*\lambda_m^*\nu_{m1}'\alpha_m\lambda_m \nu_{m1}}{1+\alpha_m\lambda_m}\right)(1+\alpha_m\lambda_m)\right)\nonumber\\[-2mm]
\overset{(c_2)}{=}&\log_2\left((1+\alpha_m\lambda_m)(1+\alpha_m^*\lambda_m^*)-\nu_{m1}'\nu_{m1}|\lambda_m|^2\right)
\nonumber\\[-2mm]
=&\log_2\left(1+|\lambda_m|^2(1-\nu_{m1}'\nu_{m1})+2\mathfrak{Re}\{\alpha_m\lambda_m\}\right),
\end{align}
where $(c_1)$ holds due to the fact that $\det({\mv{AB}})=\det({\mv{A}})\det({\mv{B}})$ and $\det({\mv{I}}_p+{\mv{CD}})=\det({\mv{I}}_q+{\mv{DC}})$ for ${\mv{C}}\in \mathbb{C}^{p\times q}$ and ${\mv{D}}\in \mathbb{C}^{q\times p}$; $(c_2)$ holds due to $|\alpha_m|^2=1$.

Based on (\ref{f_m'2}), (P1-m) is equivalent to maximizing $\mathfrak{Re}\{\alpha_m\lambda_m\}$ under the constraint in (\ref{P1mc1}) when ${\mv{A}}_m^{-1}{\mv{B}}_m$ is diagonalizable, for which we have the following proposition.
\begin{proposition}\label{prop_diagonalize}
If $\mathrm{tr}({\mv{A}}_m^{-1}{\mv{B}}_m)\neq 0$, the optimal solution to (P1-m) is given by
\vspace{-3mm}\begin{equation}
\alpha_m^{\star {\mathrm{I}}}=e^{-j\arg\{\lambda_m\}}.
\vspace{-3mm}\end{equation}
The optimal value of (P1-m) is thus given by
\vspace{-3mm}\begin{equation}
f_m^{\star {\mathrm{I}}}=\log_2\left(1+|\lambda_m|^2(1-\nu_{m1}'\nu_{m1})+2|\lambda_m|\right)+\log_2\det({\mv{A}}_m).
\vspace{-3mm}\end{equation}
\end{proposition}
\begin{IEEEproof}
Since $\mathfrak{Re}\{\alpha_m\lambda_m\}\leq |\alpha_m\lambda_m|=|\alpha_m||\lambda_m|=|\lambda_m|$, where the inequality holds with equality if and only if $\arg\{\alpha_m\}=-\arg\{\lambda_m\}$, the proof of Proposition \ref{prop_diagonalize} is thus completed.
\end{IEEEproof}
\subsubsection{Case II: Non-Diagonalizable ${\mv{A}}_m^{-1}{\mv{B}}_m$}
Next, consider the case where ${\mv{A}}_m^{-1}{\mv{B}}_m$ is non-diagonalizable. In this case, we express it as ${\mv{A}}_m^{-1}{\mv{B}}_m={\mv{u}}_m{\mv{v}}_m^H$, where ${\mv{u}}_m\in \mathbb{C}^{N_r\times 1}$, ${\mv{v}}_m\in \mathbb{C}^{N_r\times 1}$, and ${\mv{v}}_m^H{\mv{u}}_m={\mv{u}}_m^H{\mv{v}}_m=\mathrm{tr}({\mv{A}}_m^{-1}{\mv{B}}_m)=0$ according to Lemma \ref{lemma_diagonalize}. To exploit the structure of ${\mv{A}}_m^{-1}{\mv{B}}_m$ in this case, we first provide the following lemma for $\mv{u}_m$ and $\mv{v}_m$.
\begin{lemma}\label{lemma_inversion}
${\mv{I}}_{N_r}+\alpha_m{\mv{u}}_m{\mv{v}}_m^H$ is an invertible matrix, whose inversion is given by
\vspace{-3mm}\begin{equation}
({\mv{I}}_{N_r}+\alpha_m{\mv{u}}_m{\mv{v}}_m^H)^{-1}={\mv{I}}_{N_r}-\alpha_m{\mv{u}}_m{\mv{v}}_m^H.\label{inversion1}
\vspace{-3mm}\end{equation}
\end{lemma}
\begin{IEEEproof}
Lemma \ref{lemma_inversion} follows from the Sherman-Morrison-Woodbury formula \cite{Matrix_Analysis}, which states that for an invertible matrix $\mv{A}\in \mathbb{C}^{N_r\times N_r}$ and two vectors ${\mv{a}}\in \mathbb{C}^{N_r\times 1}$ and ${\mv{b}}\in \mathbb{C}^{N_r\times 1}$, ${\mv{A}}+{\mv{ab}}^H$ is invertible if and only if $1+{\mv{b}}^H{\mv{A}}^{-1}{\mv{a}}\neq 0$, and the inversion is given by $({\mv{A}}+{\mv{ab}}^H)^{-1}={\mv{A}}^{-1}-\frac{{\mv{A}}^{-1}{\mv{ab}}^H{\mv{A}^{-1}}}{1+{\mv{b}}^H{\mv{A}}^{-1}{\mv{a}}}$. Based on this, by replacing $\mv{A}$ with $\mv{I}_{N_r}$, we have $1+\alpha_m{\mv{v}}_m^H{\mv{u}}_m=1\neq 0$, thus ${\mv{I}}_{N_r}+\alpha_m{\mv{u}}_m{\mv{v}}_m^H$ is an invertible matrix, with the inversion given in (\ref{inversion1}).
\end{IEEEproof}

Based on the results in Lemma \ref{lemma_inversion}, $f_m'$ can be rewritten as
\vspace{-2mm}\begin{align}\label{f_m'3}
f_m'=&\log_2\det \left({\mv{I}}_{N_r}+\alpha_m{\mv{u}}_m{\mv{v}}_m^H+\alpha_m^*{\mv{A}}_m^{-1}{\mv{v}}_m{\mv{u}}_m^H{\mv{A}}_m\right)\nonumber\\[-2mm]
\overset{(d_1)}{=}&\log_2\det ({\mv{I}}_{N_r}+\alpha_m^*({\mv{I}}_{N_r}-\alpha_m {\mv{u}}_m{\mv{v}}_m^H){\mv{A}}_m^{-1}{\mv{v}}_m{\mv{u}}_m^H{\mv{A}}_m) + \log_2\det ({\mv{I}}_{N_r}+\alpha_m{\mv{u}}_m{\mv{v}}_m^H)\nonumber\\[-2mm]
\overset{(d_2)}{=}&\log_2\det ({\mv{I}}_{N_r}+\alpha_m^*({\mv{I}}_{N_r}-\alpha_m {\mv{u}}_m{\mv{v}}_m^H){\mv{A}}_m^{-1}{\mv{v}}_m{\mv{u}}_m^H{\mv{A}}_m)\nonumber\\[-2mm]
\overset{(d_3)}{=}&\log_2\det ({\mv{A}}_m({\mv{I}}_{N_r}+\alpha_m^*({\mv{I}}_{N_r}-\alpha_m {\mv{u}}_m{\mv{v}}_m^H){\mv{A}}_m^{-1}{\mv{v}}_m{\mv{u}}_m^H{\mv{A}}_m){\mv{A}}_m^{-1})\nonumber\\[-2mm]
\overset{(d_4)}{=}&\log_2\det ({\mv{I}}_{N_r}+\alpha_m^*{\mv{v}}_m{\mv{u}}_m^H-{\mv{A}}_m{\mv{u}}_m{\mv{v}}_m^H{\mv{A}}_m^{-1}{\mv{v}}_m{\mv{u}}_m^H)\nonumber\\[-2mm]
\overset{(d_5)}{=}&\log_2\det ({\mv{I}}_{N_r}-({\mv{I}}_{N_r}-\alpha_m^*{\mv{v}}_m{\mv{u}}_m^H){\mv{A}}_m{\mv{u}}_m{\mv{v}}_m^H{\mv{A}}_m^{-1}{\mv{v}}_m{\mv{u}}_m^H)+\log_2\det({\mv{I}}_{N_r}+\alpha_m^*{\mv{v}}_m{\mv{u}}_m^H)\nonumber\\[-2mm]
\overset{(d_6)}{=}&\log_2\det ({\mv{I}}_{N_r}-{\mv{A}}_m^{-1}{\mv{v}}_m{\mv{u}}_m^H({\mv{I}}_{N_r}-\alpha_m^*{\mv{v}}_m{\mv{u}}_m^H){\mv{A}}_m{\mv{u}}_m{\mv{v}}_m^H)\nonumber\\[-2mm]
\overset{(d_7)}{=}&\log_2\det ({\mv{I}}_{N_r}-{\mv{A}}_m^{-1}{\mv{v}}_m{\mv{u}}_m^H{\mv{A}}_m{\mv{u}}_m{\mv{v}}_m^H),
\end{align}
where $(d_1)$ can be derived in a similar manner as $(c_1)$ with the help of Lemma \ref{lemma_inversion}; $(d_2)$ holds since $\log_2\det ({\mv{I}}_{N_r}+\alpha_m{\mv{u}}_m{\mv{v}}_m^H)=\log_2\det (1+\alpha_m{\mv{v}}_m^H{\mv{u}}_m)=0$; $(d_3)$ can be derived in a similar manner as $(b_1)$ and $(b_2)$ by noting that ${\mv{A}}_m$ is invertible; $(d_4)$ holds since $|\alpha_m|^2=1$; $(d_5)$ can be derived similarly as $(d_1)$; $(d_6)$ follows from $\det({\mv{I}}_p+{\mv{CD}})=\det({\mv{I}}_q+{\mv{DC}})$ and $\log_2\det ({\mv{I}}_{N_r}+\alpha_m^*{\mv{v}}_m{\mv{u}}_m^H)=\log_2\det (1+\alpha_m^*{\mv{u}}_m^H{\mv{v}}_m)=0$; and $(d_7)$ holds since ${\mv{u}}_m^H{\mv{v}}_m=0$, and consequently ${\mv{A}}_m^{-1}{\mv{v}}_m{\mv{u}}_m^H\alpha_m^*{\mv{v}}_m{\mv{u}}_m^H{\mv{A}}_m{\mv{u}}_m{\mv{v}}_m^H$ becomes an all-zero matrix.

It is worth noting from (\ref{f_m'3}) that when ${\mv{A}}_m^{-1}{\mv{B}}_m$ is non-diagonalizable, $f_m'$ is \emph{independent} of $\alpha_m$. Therefore, we have the following proposition.
\begin{proposition}\label{prop_nondiagonalize}
	If $\mathrm{tr}({\mv{A}}_m^{-1}{\mv{B}}_m)= 0$, any $\alpha_m$ with $|\alpha_m|=1$ is an optimal solution to (P1-m). The optimal value of (P1-m) is thus given by
	\vspace{-3mm}\begin{equation}
	f_m^{\star {\mathrm{II}}}=\log_2\det ({\mv{A}}_m-{\mv{B}}_m^H{\mv{A}}_m^{-1}{\mv{B}}_m).
	\vspace{-3mm}\end{equation}
\end{proposition}
\begin{IEEEproof}
	The first half of Proposition \ref{prop_nondiagonalize} follows directly from (\ref{f_m'3}). The second half of Proposition \ref{prop_nondiagonalize} can be derived as
	$f_m^{\star {\mathrm{II}}}=\log_2\det ({\mv{I}}_{N_r}-{\mv{A}}_m^{-1}{\mv{v}}_m{\mv{u}}_m^H{\mv{A}}_m{\mv{u}}_m{\mv{v}}_m^H)+\log_2\det ({\mv{A}}_m)=\log_2\det ({\mv{A}}_m-{\mv{v}}_m{\mv{u}}_m^H{\mv{A}}_m{\mv{u}}_m{\mv{v}}_m^H)
	=\log_2\det ({\mv{A}}_m-{\mv{B}}_m^H{\mv{A}}_m^{-1}{\mv{B}}_m)$, by noting that ${\mv{A}}_m^{-1}{\mv{B}}_m={\mv{u}}_m{\mv{v}}_m^H$ holds. This thus completes the proof of Proposition \ref{prop_nondiagonalize}.
\end{IEEEproof}

Based on the results in Proposition \ref{prop_nondiagonalize}, we set $\alpha_m^{\star {\mathrm{II}}}=1$ as the optimal solution to (P1-m) in this case without loss of optimality.
\subsubsection{Summary of the Optimal Solution to Problem (P1-m)}
To summarize, the optimal solution to (P1-m) is given by
\vspace{-2mm}\begin{equation}\label{alpham}
\alpha_m^\star=\begin{cases}
e^{-j\arg\{\lambda_m\}},\ &\mathrm{if}\ \mathrm{tr}({\mv{A}}_m^{-1}{\mv{B}}_m)\neq 0\\[-2mm]
1,\ &\mathrm{otherwise}.
\end{cases}
\vspace{-2mm}\end{equation}
The corresponding optimal value of (P1-m) is given by
\vspace{-3mm}\begin{equation}\label{fm}
f_m^\star=\begin{cases}
\log_2\left(1+|\lambda_m|^2(1-\nu_{m1}'\nu_{m1})+2|\lambda_m|\right)+\log_2\det({\mv{A}}_m),\ &\mathrm{if}\ \mathrm{tr}({\mv{A}}_m^{-1}{\mv{B}}_m)\neq 0\\[-2mm]
\log_2\det ({\mv{A}}_m-{\mv{B}}_m^H{\mv{A}}_m^{-1}{\mv{B}}_m),\ &\mathrm{otherwise}.
\end{cases}
\vspace{-3mm}\end{equation}
\vspace{-9mm}
\subsection{Overall Algorithm}
\vspace{-2mm}
With the optimal solution to (P1-m) derived above, we are ready to complete our proposed alternating optimization algorithm for solving (P1). Specifically, we first randomly generate $L> 1$ sets of $\{\alpha_m\}_{m=1}^M$ with $|\alpha_m|=1,\forall m$ and phases of $\alpha_m$'s following the uniform distribution in $[0,2\pi)$. By obtaining the optimal transmit covariance matrix $\mv{Q}$ for each set of $\{\alpha_m\}_{m=1}^M$ according to (\ref{Qstar}) as well as the corresponding channel capacity, we select the set with maximum capacity as the initial point. The algorithm then proceeds by iteratively solving the two subproblems presented in Section \ref{sec_framework}, until convergence is reached. The overall algorithm is summarized in Algorithm \ref{algo1}.
\vspace{-3mm}
\begin{algorithm}[h]\label{algo1}
	\caption{Proposed Algorithm for Problem (P1)}
	\SetKwData{Index}{Index}
	\KwIn{$\mv{H}$, ${\mv{R}}$, ${\mv{T}}$, $P$, $\sigma^2$, $L$}
	\KwOut{$\mv{\phi}$, ${\mv{Q}}$}
	Randomly generate $L$ independent realizations of $\{{\alpha}_m\}_{m=1}^M$, and obtain the optimal transmit covariance matrix $\mv{Q}$ according to (\ref{Qstar}) for each realization.\\
	Select $\{\tilde{\alpha}_m^\star\}_{m=1}^M$ and the corresponding $\tilde{\mv{Q}}^\star$ as the realization yielding the largest objective value of (P1).\\
	Initialize $\alpha_m=\tilde{\alpha}_m^\star,\ m=1,...,M$; $\mv{Q}=\tilde{\mv{Q}}^\star$.\\
	\For{$m=1\rightarrow M$}{
	Obtain $\mv{A}_m$ and $\mv{B}_m$ according to (\ref{A_mB_m}).\\
	Obtain the optimal solution to (P1-m) according to (\ref{alpham}).
}
	Obtain the optimal solution of ${\mv{Q}}$ to (P1) with given $\{{\alpha}_m\}_{m=1}^M$ according to (\ref{Qstar}).\\
	Check convergence. If yes, stop; if not, go to Step 4.\\
	Set ${\mv{\phi}}=\mathrm{diag}\{\alpha_1,...,\alpha_M\}$.
\end{algorithm}
\vspace{-6mm}

Note that in Algorithm \ref{algo1}, we have obtained the {\emph{optimal}} solution to every subproblem. Therefore, \emph{monotonic convergence} of Algorithm \ref{algo1} is guaranteed, since the algorithm yields non-decreasing objective value of (P1) over the iterations, which is also upper-bounded by a finite capacity. Moreover, since the objective function of (P1) is differentiable and all the variables $\{\alpha_m\}_{m=1}^M$ and $\mv{Q}$ are not coupled in the constraints, any limit point of the iterations generated by Algorithm \ref{algo1} satisfies the Karush-Kuhn-Tucker (KKT) condition of (P1) \cite{Convergence}. By further setting the convergence criteria of Algorithm \ref{algo1} as that the objective function of (P1) cannot be further increased by optimizing any variable in $\{\alpha_m\}_{m=1}^M\cup\mv{Q}$, Algorithm \ref{algo1} is guaranteed to converge to at least a \emph{locally optimal} solution of (P1). Finally, it is worth noting that the (worst-case) complexity for Algorithm \ref{algo1} can be shown to be $\mathcal{O}(N_rN_t(M+\min(N_r,N_t))L+((3N_r^3+2N_r^2N_t+N_t^2)M+N_rN_t\min(N_r,N_t))I)$ with $I$ denoting the number of outer iterations (i.e., the number of times that Steps 4--8 are repeated), which is polynomial over $N_r$, $N_t$, and $M$.
\vspace{-2mm}
\subsection{Alternative Solutions to Problem (P1) in Low-/High-SNR Regimes}\label{sec_alternative}
\vspace{-1mm}
In the previous subsections, we have proposed an alternating optimization algorithm that can handle the general MIMO channel capacity maximization problem (P1). In this subsection, we consider the MIMO channel under either asymptotically \emph{low-SNR} regime or asymptotically \emph{high-SNR} regime, and derive their corresponding channel capacities in more tractable forms in terms of the reflection coefficients, based on which two low-complexity alternative solutions to (P1) are proposed, respectively.
\subsubsection{Low-SNR Regime (Strongest Eigenchannel Power Maximization)}\label{sec_alternativelow}
First, we consider the low-SNR regime, which may correspond to the case with low transmission power (e.g., uplink) and/or long distance between the transmitter and the receiver. In this regime, the optimal transmission strategy is beamforming over the \emph{strongest eigenmode} of the effective MIMO channel, $\tilde{\mv{H}}$, by allocating all transmit power to the strongest eigenchannel \cite{Fundamental}. Specifically, the optimal transmit covariance matrix that maximizes the capacity is given by $\mv{Q}=P\tilde{\mv{v}}_1\tilde{\mv{v}}_1^H$, where ${\tilde{\mv{v}}}_1\in \mathbb{C}^{N_t\times 1}$ denotes the strongest right singular vector of $\tilde{\mv{H}}$. The capacity in (\ref{capacity}) can be thus rewritten as
\vspace{-3mm}\begin{equation}\label{capacity_lowSNR}
C_{\mathrm{L}}=\log_2\left(1+P[\tilde{\mv{\Lambda}}]_{\max}^2/\sigma^2\right)=\underset{\|\bar{\mv{x}}\|=1,\|\bar{\mv{y}}\|=1}{\max}\log_2\left(1+P|\bar{\mv{x}}^H\tilde{\mv{H}}\bar{\mv{y}}|^2/\sigma^2\right),
\vspace{-3mm}\end{equation}
where $[\tilde{\mv{\Lambda}}]_{\max}\overset{\Delta}{=}\underset{i=1,...,D}{\max}\ [\tilde{\mv{\Lambda}}]_{i,i}=\underset{\|\bar{\mv{x}}\|=1,\|\bar{\mv{y}}\|=1}{\max}|\bar{\mv{x}}^H\tilde{\mv{H}}\bar{\mv{y}}|$ denotes the strongest singular value of $\tilde{\mv{H}}$, or the strongest eigenchannel gain; $\bar{\mv{x}}\in \mathbb{C}^{N_r\times 1}$ and $\bar{\mv{y}}\in \mathbb{C}^{N_t\times 1}$. Based on (\ref{capacity_lowSNR}), the capacity maximization problem in the low-SNR regime can be solved by maximizing the \emph{strongest eigenchannel power} of $\tilde{\mv{H}}$, $[\tilde{\mv{\Lambda}}]_{\max}^2$, by jointly optimizing the reflection coefficients $\{\alpha_m\}_{m=1}^M$ and the auxiliary vectors $\bar{\mv{x}}$ and $\bar{\mv{y}}$, which is a non-convex optimization problem since $[\tilde{\mv{\Lambda}}]_{\max}^2$ can be shown to be a non-concave function over $\{\alpha_m\}_{m=1}^M$. Note that with given $\{\alpha_m\}_{m=1}^M$, the optimal solutions to $\bar{\mv{x}}$ and $\bar{\mv{y}}$ can be shown to be the strongest left and right singular vectors of $\tilde{\mv{H}}$, respectively. On the other hand, with given $\bar{\mv{x}}$ and $\bar{\mv{y}}$, the optimal $\{\alpha_m\}_{m=1}^M$ that maximizes $|\bar{\mv{x}}^H\tilde{\mv{H}}\bar{\mv{y}}|^2=|\bar{\mv{x}}^H{\mv{H}}\bar{\mv{y}}+\bar{\mv{x}}^H{\mv{R\phi T}}\bar{\mv{y}}|^2=|\bar{\mv{x}}^H{\mv{H}}\bar{\mv{y}}+\sum_{m=1}^M\alpha_m[\bar{\mv{x}}^H{\mv{R}}]_m[{\mv{T}}\bar{\mv{y}}]_m|^2$ can be easily shown to be $\alpha_m^\star=e^{j(\arg\{\bar{\mv{x}}^H{\mv{H}}\bar{\mv{y}}\} - \arg\{[\bar{\mv{x}}^H{\mv{R}}]_m[{\mv{T}}\bar{\mv{y}}]_m\})}$, $\forall m\in \mathcal{M}$. Therefore, by a similar alternating optimization as Algorithm \ref{algo1}, a locally optimal solution to the capacity maximization problem in the low-SNR regime can be obtained via iteratively optimizing the two sets of variables $\{\alpha_m\}_{m=1}^M$ and $\{\bar{\mv{x}},\bar{\mv{y}}\}$ with the other set being fixed at each time. Note that similar to Algorithm \ref{algo1}, an initial point of the algorithm can be found by randomly generating $L> 1$ sets of $\{\alpha_m\}_{m=1}^M$ and selecting the set with the largest $[\tilde{\mv{\Lambda}}]_{\max}^2$. The required complexity for the overall algorithm can be shown to be $\mathcal{O}(N_rN_t(M+\min(N_r,N_t))L+N_rN_t(M+\min(N_r,N_t))I)$, which is generally lower than that of Algorithm \ref{algo1} since there is no need to compute $\mv{A}_m^{-1}{\mv{B}}_m$'s and their EVDs as in Algorithm \ref{algo1}, with $I$ denoting the number of outer iterations.
\subsubsection{High-SNR Regime (Channel Total Power Maximization)}\label{sec_alternativehigh}
Next, we consider the high-SNR regime, which may correspond to the case with high transmission power (e.g., downlink) and/or short distance between the transmitter and receiver. In this regime, it is asymptotically optimal to allocate equal power among all available eigenmodes \cite{Fundamental}, and the channel capacity in (\ref{capacity}) can be approximated as
\vspace{-3mm}\begin{equation}\label{capacity_highSNR}
C_{\mathrm{H}}\!\approx\! \sum_{i=1}^D\log_2\left(1+\frac{P[\tilde{\mv{\Lambda}}]_{i,i}^2}{D\sigma^2}\right)\!\leq\! D\log_2\left(1+\frac{P\sum_{i=1}^D [\tilde{\mv{\Lambda}}]_{i,i}^2}{D^2\sigma^2}\right)\!=\!D\log_2\left(1+\frac{P\|\tilde{\mv{H}}\|_F^2}{D^2\sigma^2}\right),
\vspace{-3mm}\end{equation}
where the inequality holds with equality if and only if all eigenchannel powers $[\tilde{\mv{\Lambda}}]_{i,i}$'s are equal \cite{Fundamental}. Motivated by (\ref{capacity_highSNR}), we propose to find an approximate solution to (P1) in the high-SNR regime by maximizing the \emph{channel total power} via optimization of $\mv{\phi}$ subject to the constraints in (\ref{P1c1}) and (\ref{P1c2}), for which the problem is reformulated as
\vspace{-3mm}\begin{equation}
\mbox{(P-Power)}\quad \underset{\mv{\phi}:(\ref{P1c1}),(\ref{P1c2})}{\max}\ \|\tilde{\mv{H}}\|_F^2.
\vspace{-3mm}\end{equation} 
Note that (P-Power) is a non-convex optimization problem since $\|\tilde{\mv{H}}\|_F^2$ can be shown to be a non-concave function over $\mv{\phi}$. In the following, we find a high-quality suboptimal solution to it via alternating optimization based on the expression of $\tilde{\mv{H}}$ in (\ref{effchannel}). Specifically, we first express the channel total power as
\vspace{-3mm}\begin{align}\label{channelpower}
&\!\!\|\tilde{\mv{H}}\|_F^2\!=\!\mathrm{tr}(\tilde{\mv{H}}\tilde{\mv{H}}^H)\!=\!\mathrm{tr}\left(\left(\mv{H}\!+\!\sum_{i=1,i\neq m}^M \alpha_i \mv{r}_i\mv{t}_i^H\!+\!\alpha_m\mv{r}_m\mv{t}_m^H\right)\left(\mv{H}\!+\!\sum_{i=1,i\neq m}^M \alpha_i \mv{r}_i\mv{t}_i^H\!+\!\alpha_m\mv{r}_m\mv{t}_m^H\right)^H\right)
\nonumber\\[-2mm]
&\!\!\!\overset{(e)}{=}\!\left\|\mv{H}\!\!+\!\!\!\!\!\sum_{i=1,i\neq m}^M \!\!\!\alpha_i \mv{r}_i\mv{t}_i^H\right\|_F^2\!\!\!+\!2\mathfrak{Re}\!\left\{\!\alpha_m^*\mv{r}_m^H\!\left(\!\mv{H}\!\!+\!\!\!\!\!\sum_{i=1,i\neq m}^M\!\!\!\!\alpha_i\mv{r}_i\mv{t}_i^H\!\right)\!{\mv{t}}_m\!\right\}\!+\!\mathrm{tr}(\mv{r}_m\mv{t}_m^H\mv{t}_m{\mv{r}}_m^H),\quad \forall m\!\in\! \mathcal{M},\!\!\!\!
\end{align}
where $(e)$ follows from $|\alpha_m|^2=1,\forall m$ and $\mathrm{tr}(\mv{AB})=\mathrm{tr}(\mv{BA})$ for two equal-sized square matrices $\mv{A}$ and $\mv{B}$. Hence, we have the following proposition for any $m\in \mathcal{M}$.
\begin{proposition}\label{prop_Power}
	With any given $\{\alpha_i,i\neq m\}_{i=1}^M$, the optimal $\alpha_m$ to (P-Power) is given by
	\vspace{-3mm}
	\begin{equation}\label{alphaoptPower}
	\alpha_m^\star=e^{j\arg\left\{{\mv{r}}_m^H\left({\mv{H}}+\sum_{i=1,i\neq m}^M \alpha_i{\mv{r}}_i{\mv{t}}_i^H\right){\mv{t}}_m\right\}}.
	\vspace{-3mm}
	\end{equation}
	The corresponding optimal value of (P-Power) is 
	$\|\mv{H}+\sum_{i=1,i\neq m}^M \alpha_i \mv{r}_i\mv{t}_i^H\|_F^2+\mathrm{tr}(\mv{r}_m\mv{t}_m^H\mv{t}_m\mv{r}_m^H)+2|{\mv{r}}_m^H(\mv{H}+\sum_{i=1,i\neq m}^M\alpha_i\mv{r}_i\mv{t}_i^H)\mv{t}_m|$.
\end{proposition}
\begin{IEEEproof}
	Let $\beta_m={\mv{r}}_m^H({\mv{H}}+\sum_{i=1,i\neq m}^M \alpha_i{\mv{r}}_i{\mv{t}}_i^H){\mv{t}}_m$. Since $\mathfrak{Re}\{\alpha_m^*\beta_m\}\leq |\alpha_m^*\beta_m|=|\beta_m|$, where the inequality holds with equality if and only if $\alpha_m=\arg\{\beta_m\}$, Proposition \ref{prop_Power} thus holds.
\end{IEEEproof}
Therefore, a locally optimal solution to (P-Power) can be obtained via alternating optimization by iteratively optimizing one reflection coefficient $\alpha_m$ with $\{\alpha_i,i\neq m\}_{i=1}^M$ being fixed at each time, where an initial point of the algorithm can be found via a similar approach as Algorithm \ref{algo1} by randomly generating $L$ sets of $\{\alpha_m\}_{m=1}^M$ and selecting the one with the largest channel total power. This algorithm can be shown to require complexity $\mathcal{O}(N_tN_r(M+N_r)L+N_tN_rMI)$, which is also lower than that of Algorithm \ref{algo1} in general, \hbox{where $I$ denotes the number of outer iterations.}
\vspace{-1mm}
\subsection{Solution to Problem (P1) with Single-Antenna Transmitter/Receiver}\label{sec_MISO}
\vspace{-1mm}
So far, we have investigated (P1) for the general MIMO channel with $N_r\geq 1$ and $N_t\geq 1$, by considering parallel transmissions of multiple data streams in general. In this subsection, we study (P1) for the special cases with $N_r=1$ or $N_t=1$, where only one data stream can be transmitted. This leads to more simplified expressions of the optimal transmit covariance matrix as well as the channel capacity, based on which we propose simpler alternating optimization algorithms for solving (P1) that require much lower complexity compared to Algorithm \ref{algo1} in the case of $N_r=1$ or $N_t=1$.

First, we consider (P1) for the MISO case with $N_t\geq 1$ and $N_r=1$, where the channel matrices $\mv{H}\in \mathbb{C}^{N_r\times N_t}$ and $\mv{R}\in \mathbb{C}^{N_r\times M}$ can be rewritten as ${\mv{h}}^H\in \mathbb{C}^{1\times N_t}$ and ${\mv{r}}^H=[r_1^*,...,r_M^*]\in \mathbb{C}^{1\times M}$, respectively, and the overall effective channel can be expressed as $\tilde{\mv{h}}^H={\mv{h}}^H+{\mv{r}}^H\mv{\phi}{\mv{T}}\in \mathbb{C}^{1\times N_t}$. 
Note that in this case, the optimal transmit covariance matrix is given by the maximum ratio transmission (MRT) \cite{Fundamental}, namely, $\mv{Q}^\star=P\tilde{\mv{h}}\tilde{\mv{h}}^H/\|\tilde{\mv{h}}\|^2$. Consequently, the MISO channel capacity can be rewritten as $C_{\mathrm{MISO}}=\log_2 (1+\tilde{\mv{h}}^H{\mv{Q}}^\star\tilde{\mv{h}}/\sigma^2)=\log_2 (1+P\|\tilde{\mv{h}}\|^2/\sigma^2)$, which is an explicit function of the effective channel $\tilde{\mv{h}}^H$. Thus, (P1) can be equivalently transformed into the following problem for maximizing the channel total power via optimizing $\mv{\phi}$:
\vspace{-3mm}
\begin{equation}
\mbox{(P1-MISO)}\quad \underset{\mv{\phi}:(\ref{P1c1}),(\ref{P1c2})}{\max}\  \big\|{\mv{h}}^H+\sum_{i=1}^M \alpha_i r_i^*{\mv{t}}_i^H\big\|^2.
\vspace{-3mm}
\end{equation}
Note that (P1-MISO) is in fact a degenerated version of (P-Power) in Section \ref{sec_alternativehigh} for maximizing the MIMO channel total power, thus can be handled via a similar approach. Specifically, the following proposition follows directly from Proposition \ref{prop_Power}.
\begin{proposition}\label{prop_MISO}
	With any given $\{\alpha_i,i\neq m\}_{i=1}^M$, the optimal $\alpha_m$ to (P1-MISO) is given by
	\vspace{-3mm}\begin{equation}\label{alphaoptMISO}
	\alpha_m^\star=e^{j\arg\left\{r_m\left({\mv{h}}^H+\sum_{i=1,i\neq m}^M \alpha_ir_i^*{\mv{t}}_i^H\right){\mv{t}}_m\right\}}.
	\vspace{-3mm}\end{equation}
	The corresponding optimal value of (P1-MISO) is
	$\|{\mv{h}}^H+\sum_{i=1,i\neq m}^M \alpha_i r_i^*{\mv{t}}_i^H\|^2+\|r_m^*{\mv{t}}_m^H\|^2+2|r_m({\mv{h}}^H+\sum_{i=1,i\neq m}^M \alpha_ir_i^*{\mv{t}}_i^H){\mv{t}}_m|$.
\end{proposition}

Based on Proposition \ref{prop_MISO}, the proposed alternating optimization algorithm for (P-Power) can be readily applied for solving (P1-MISO), by successively optimizing each reflection coefficient $\alpha_m$ with the other $M-1$ ones being fixed at each time, which is guaranteed to converge to at least a \emph{locally optimal} solution to (P1-MISO) with complexity $\mathcal{O}(N_tML+N_tMI)$, with $L$ and $I$ denoting the numbers of initializations and outer iterations, respectively. It is worth noting that the complexity of this algorithm is generally lower than that of Algorithm \ref{algo1} with $N_r=1$, since the MISO channel capacity can be explicitly expressed as a function of  $\{\alpha_m\}_{m=1}^M$, thus eliminating the need of iteratively solving $\mv{Q}$ in the alternating optimization.

Next, we consider a SIMO system with $N_r\geq 1$ and $N_t=1$, where the channel matrices $\mv{H}$ and $\mv{T}$ can be rewritten as ${\mv{h}}\in \mathbb{C}^{N_r\times 1}$ and $\mv{t}\in \mathbb{C}^{M\times 1}$, respectively, and the overall effective SIMO channel is given by $\tilde{\mv{h}}=\mv{h}+{\mv{R}}\mv{\phi}{\mv{t}}\in \mathbb{C}^{N_r\times 1}$. Note that with $N_t=1$, the optimal transmit covariance matrix can be easily shown to be ${\mv{Q}}^\star=P$. Therefore, the corresponding channel capacity is given by $C_{\mathrm{SIMO}}=\log_2\det({\mv{I}}_{N_r}+\tilde{\mv{h}}{\mv{Q}}^\star\tilde{\mv{h}}^H/\sigma^2)=\log_2(1+P\|\tilde{\mv{h}}\|^2/\sigma^2)$. Notice that the above SIMO channel capacity is in a similar form as the MISO channel capacity $C_{\mathrm{MISO}}$, which can be maximized by maximizing the channel total power. Hence, the proposed alternating optimization algorithm for the MISO case can be \hbox{also applied to the SIMO case.}

\vspace{-3mm}
\section{Capacity Maximization for MIMO-OFDM System}\label{sec_OFDM}
\vspace{-2mm}
In this section, we extend our results on the narrowband transmission to the broadband MIMO-OFDM systems under frequency-selective channels. Let $L_{\mathrm{D}}$, $L_{\mathrm{TI}}$, and $L_{\mathrm{IR}}$ denote the numbers of delayed taps in the time-domain impulse responses for the direct link, the transmitter-IRS link, and the IRS-receiver link, respectively, and let $\bar{\mv{H}}_l\in \mathbb{C}^{N_r\times N_t},\ l\in \{0,...,L_{\mathrm{D}}-1\}$, $\bar{\mv{T}}_l\in \mathbb{C}^{M\times N_t},\ l\in \{0,...,L_{\mathrm{TI}}-1\}$, $\bar{\mv{R}}_l\in \mathbb{C}^{N_r\times M},\ l\in \{0,...,L_{\mathrm{IR}}-1\}$ denote the corresponding time-domain channel matrices at each $l$th tap, respectively. Note that the overall impulse response of the reflected link is the convolution of the transmitter-IRS channel $\{\bar{\mv{T}}_l\}_{l=0}^{L_{\mathrm{TI}}-1}$, the IRS reflection matrix $\mv{\phi}$, and the IRS-receiver channel $\{\bar{\mv{R}}_l\}_{l=0}^{L_{\mathrm{IR}}-1}$. Thus, the overall impulse response from the transmitter to the receiver consists of at most $L_{\max}=\max\{L_{\mathrm{D}},L_{\mathrm{TI}}+L_{\mathrm{IR}}-1\}$ delayed taps, and the overall time-domain effective channel at each $l$th tap can be expressed as $\tilde{\bar{\mv{H}}}_l=\bar{\mv{H}}_l+\sum_{q=0}^{L_{\mathrm{IR}}-1}\bar{\mv{R}}_q\mv{\phi}\bar{\mv{T}}_{l-q},\ l=0,...,L_{\max}-1$, where we define $\bar{\mv{H}}_l={\mv{0}},\ l\in \{L_{\mathrm{D}},...,L_{\max}-1\}$, and $\bar{\mv{T}}_l={\mv{0}},\ l\in \{1-L_{\mathrm{IR}},...,-1\}\cup\{L_{\mathrm{TI}},...,L_{\max}-1\}$. We consider an OFDM system with $N_f>1$ frequency subcarriers in total, among which $N$ subcarriers are allocated for our considered point-to-point transmission, with $1< N\leq N_f$. Therefore, in the frequency-domain, the channel matrix for the direct link at each $n$th subcarrier is given by ${\mv{H}}[n]=\sum_{l=0}^{L_{\mathrm{D}}-1}\bar{\mv{H}}_le^{-j2\pi(n-1)l/N},\ n=1,...,N$. The frequency-domain channel matrices for the transmitter-IRS link and the IRS-receiver link can be similarly obtained and denoted as $\{{\mv{T}}[n]\}_{n=1}^{N}$ and $\{{\mv{R}}[n]\}_{n=1}^{N}$, respectively. Based on the convolution theorem, the overall effective channel from the transmitter to the receiver in the frequency-domain can be expressed as
\vspace{-3mm}\begin{equation}
\tilde{\mv{H}}[n]={\mv{H}}[n]+{\mv{R}}[n]{\mv{\phi}}{\mv{T}}[n]={\mv{H}}[n]+\sum_{m=1}^M \alpha_m{\mv{r}}_m[n]{\mv{t}}_m[n]^H
,\quad n=1,...,N,
\vspace{-3mm}\end{equation}
where we denote $\mv{R}[n]=[{\mv{r}}_1[n],...,{\mv{r}}_M[n]],\forall n$ and $\mv{T}[n]=[{\mv{t}}_1[n],...,{\mv{t}}_M[n]]^H,\forall n$.

We aim to maximize the capacity of the above MIMO-OFDM system by jointly optimizing the transmit covariance matrices (each for a different subcarrier) and the IRS reflection coefficients (common for all the subcarriers). Specifically, an \emph{individual} transmit covariance matrix denoted by ${\mv{Q}}[n]\in \mathbb{C}^{N_t\times N_t}$ with ${\mv{Q}}[n]\succeq {\mv{0}}$ is designed for each subcarrier $n$; while in contrast, only \emph{one} set of reflection coefficients $\{\alpha_m\}_{m=1}^M$ is designed for data transmissions at all $N$ subcarriers due to the lack of baseband processing and thus ``frequency-selective'' passive beamforming capabilities at the IRS, which is the main new consideration as compared to the previous case with frequency-flat channels. We consider an average transmit power constraint over all subcarriers given by $\frac{1}{N}\sum_{n=1}^N \mathrm{tr}({\mv{Q}}[n])\leq P$, and let $\bar{\sigma}^2$ denote the average noise power at each subcarrier. The capacity for our considered MIMO-OFDM transmission is thus given by
\vspace{-3mm}\begin{equation}\label{capacity_OFDM}
C_{\mathrm{OFDM}}=\underset{\scriptstyle \{\small{\mv{Q}}[n]\}_{n=1}^N:\small{\mv{Q}}[n]\succeq \mv{0},\forall n\atop \scriptstyle \frac{1}{N}\sum_{n=1}^N\mathrm{tr}(\small{\mv{Q}}[n])\leq P}{\max}\frac{N_f}{N_f+\mu}\frac{1}{N}\sum_{n=1}^N \log_2\det \left({\mv{I}}_{N_r}+\frac{1}{\bar{\sigma}^2}\tilde{\mv{H}}[n]{\mv{Q}}[n]\tilde{\mv{H}}[n]^H\right),
\vspace{-3mm}\end{equation}
where $\mu\geq L_{\max}$ denotes the cyclic prefix length of the OFDM system. By dropping the constant term in $C_{\mathrm{OFDM}}$, the optimization problem is thus formulated as
\vspace{-2mm}
\begin{align}
\mbox{(P2)}\quad \underset{\mv{\phi},\{{\mv{Q}}[n]\}_{n=1}^N}{\max}\  &\sum_{n=1}^N \log_2\det \left({\mv{I}}_{N_r}+\frac{1}{\bar{\sigma}^2}\tilde{\mv{H}}[n]{\mv{Q}}[n]\tilde{\mv{H}}[n]^H\right)\\[-2mm]
\mbox{s.t.}\quad & {\mv{\phi}}=\mathrm{diag}\{\alpha_1,...,\alpha_M\}\label{P2c1}\\[-2mm]
&|\alpha_m|= 1,\quad m=1,...,M\label{P2c2}\\[-2mm]
&\frac{1}{N}\sum_{n=1}^N\mathrm{tr}(\mv{Q}[n])\leq P\label{P2c3}\\[-2mm]
& {\mv{Q}}[n] \succeq {\mv{0}},\quad n=1,...,N.\label{P2c4}
\end{align}
It is worth noting that (P2) is more challenging to solve as compared to (P1) in the frequency-flat channel case, and our proposed alternating optimization algorithm in Section \ref{sec_solution} cannot be directly applied for solving (P2), due to the following reasons. First, the objective function of (P2) is the summation of $N>1$ logarithm determinant functions, which is a non-concave function over $\{\alpha_m\}_{m=1}^M$ and also more complicated than (\ref{P1obj}) in the narrowband case. This thus makes it difficult to derive the optimal solution to each reflection coefficient $\alpha_m$ with given $\{{\mv{Q}}[n]\}_{n=1}^N$ and $\{\alpha_i,i\neq m\}_{i=1}^M$ in closed-form. Moreover, due to the uni-modular constraint on each $\alpha_m$ in (\ref{P2c2}), (P2) is a non-convex optimization problem over each $\alpha_m$ with given $\{{\mv{Q}}[n]\}_{n=1}^N$ and $\{\alpha_i,i\neq m\}_{i=1}^M$, thus making it also difficult to obtain the optimal $\alpha_m$ even numerically via standard convex optimization techniques. In the following, we tackle the above difficulties by applying the \emph{convex relaxation} technique, and propose a new alternating optimization algorithm for solving (P2) via iteratively optimizing $\{{\mv{Q}}[n]\}_{n=1}^N$ and each $\alpha_m$.

To start with, we first transform the objective function of (P2) denoted as $f_{\mathrm{OFDM}}$ into a more tractable equivalent form over $\{\alpha_m\}_{m=1}^M$ by leveraging $|\alpha_m|^2= 1,\forall m$:
\vspace{-2mm}\begin{align}
&f_{\mathrm{OFDM}}\!=\!\sum_{n=1}^N\log_2\mathrm{det}\Bigg({\mv{I}}_{N_r}\!+\!\frac{1}{\bar{\sigma}^2}\bigg(\mv{H}[n]\mv{Q}[n]\mv{H}[n]^H\!+\!\sum_{i=1}^M\sum_{j=1,j\neq i}^M\alpha_i\alpha_j^*{\mv{r}}_i[n]{\mv{t}}_i[n]^{H}\mv{Q}[n]{\mv{t}}_j[n]{\mv{r}}_j[n]^H\nonumber\\[-2mm]
&+\!\sum_{i=1}^M \! \mv{r}_i[n]\mv{t}_i[n]^H\mv{Q}[n]\mv{t}_i[n]\mv{r}_i[n]^H\!
\!+\!\sum_{i=1}^M\alpha_i^*{\mv{H}}\mv{Q}[n]{\mv{t}}_i[n]{\mv{r}}_i[n]^H\!\!+\!\alpha_i{\mv{r}}_i[n]{\mv{t}}_i[n]^{H}\mv{Q}[n]{\mv{H}}[n]^{H}\!\bigg)\!\Bigg).\!\!\label{C_OFDM2}
\end{align}
Then, we propose to relax the constraints in (\ref{P2c2}) into convex constraints given by $|\alpha_m|\leq 1,m=1,...,M$, so as to find an approximate solution to (P2) by solving the following problem under the relaxed constraints: 
\vspace{-3mm}\begin{equation}
\mbox{(P2')}\quad \underset{\scriptstyle {\small{\mv{\phi}}},\{{\small{\mv{Q}}}[n]\}_{n=1}^N:(\ref{P2c1}),(\ref{P2c3}),(\ref{P2c4})\atop \scriptstyle |\alpha_m|\leq 1,\ m=1,...,M}{\max}\  f_{\mathrm{OFDM}}.
\vspace{-3mm}\end{equation}
Note that with given $\{\mv{Q}[n]\}_{n=1}^N$ and any $M-1$ reflection coefficients $\{\alpha_i,i\neq m\}_{i=1}^M$, (\ref{C_OFDM2}) is a concave function over the remaining reflection coefficient $\alpha_m$. Thus, (P2') is a convex optimization problem over $\alpha_m$, for which the optimal solution can be obtained via existing software, e.g., CVX \cite{cvx}. On the other hand, we further rewrite (\ref{C_OFDM2}) {\hbox{as the following equivalent form:}}
\vspace{-2mm}\begin{equation}
\!\!f_{\mathrm{OFDM}}\!=\!\!\sum_{n=1}^N\log_2\mathrm{det}\!\left(\!\!{\mv{I}}_{N_r}\!\!+\!\frac{1}{\bar{\sigma}^2}\!\left(\!\tilde{\mv{H}}[n]\mv{Q}[n]\tilde{\mv{H}}[n]^H\!\!+\!\!\sum_{m=1}^M(1\!-\!|\alpha_m|^2){\mv{r}}_m[n]{\mv{t}}_m[n]^H\mv{Q}[n]\mv{t}_m[n]\mv{r}_m[n]^H\!\right)\!\right),\!\label{C_OFDM3}
\vspace{-2mm}\end{equation}
which can be shown to be a concave function of $\{{\mv{Q}}[n]\}_{n=1}^N$. Thus, (P2') is a convex optimization problem over the transmit covariance matrices $\{{\mv{Q}}[n]\}_{n=1}^N$ with given $\{\alpha_m\}_{m=1}^M$, and the optimal $\{{\mv{Q}}[n]\}_{n=1}^N$ can be obtained via CVX.\footnote{It is worth noting that if $|\alpha_m|=1,\forall m$ holds, the optimal $\{{\mv{Q}}[n]\}_{n=1}^N$ that maximizes $f_{\mathrm{OFDM}}$ as well as the original objective function of (P2) can be obtained in closed-form by per-subcarrier eigenmode transmission and joint space-frequency water-filling \cite{Fundamental}. However, with given $\{\alpha_m\}_{m=1}^M$ under the relaxed constraints $|\alpha_m|\leq 1$ which may not satisfy $|\alpha_m|=1,\forall m$, a closed-form optimal solution of $\{{\mv{Q}}[n]\}_{n=1}^N$ for maximizing $f_{\mathrm{OFDM}}$ in (\ref{C_OFDM3}) is generally unknown, to our best knowledge.} Therefore, the alternating optimization framework proposed in Section \ref{sec_solution} can be similarly applied for solving (P2') by iteratively optimizing the transmit covariance matrix set $\{{\mv{Q}}[n]\}_{n=1}^N$ and one of the reflection coefficients in $\{\alpha_m\}_{m=1}^M$, which is guaranteed to converge to at least a locally optimal solution to (P2'). If all the obtained reflection coefficients for (P2') denoted by $\{{\alpha}_m\}_{m=1}^M$ satisfy the constraints in (\ref{P2c2}), the relaxation is tight, and $\{{\alpha}_m\}_{m=1}^M$ is also a locally optimal solution to (P2). Otherwise, an approximate solution to (P2) can be obtained by normalizing the amplitudes of the obtained $\{\alpha_m\}_{m=1}^M$ to one and computing the optimal $\{{\mv{Q}}[n]\}_{n=1}^N$ based on the normalized $\{\alpha_m\}_{m=1}^M$. Note that the initialization method in Algorithm \ref{algo1} can be similarly applied in this case. The overall algorithm is summarized in Algorithm \ref{algo_OFDM}.
\vspace{-4mm}
\begin{algorithm}[h]\label{algo_OFDM}
	\caption{Proposed Algorithm for Problem (P2)}
	\SetKwData{Index}{Index}
	\KwIn{$\{\mv{H}[n]\}_{n=1}^N$, $\{{\mv{R}}[n]\}_{n=1}^N$, $\{{\mv{T}}[n]\}_{n=1}^N$, $P$, $\sigma^2$, $L$}
	\KwOut{$\mv{\phi}$,$\{{\mv{Q}}[n]\}_{n=1}^N$}
	Randomly generate $L$ independent realizations of $\{{\alpha}_m\}_{m=1}^M$, and obtain the optimal transmit covariance matrices $\{{\mv{Q}}[n]\}_{n=1}^N$ via CVX.\\
	Select $\{\tilde{\alpha}_m^\star\}_{m=1}^M$ and the corresponding $\{{\mv{Q}}^\star[n]\}_{n=1}^N$ as the realization yielding the largest objective value of (P2).\\
	Initialize $\alpha_m=\tilde{\alpha}_m^\star,\ m=1,...,M$; $\mv{Q}[n]=\tilde{\mv{Q}}^\star[n],\ n=1,...,N$.\\
	\For{$m=1\rightarrow M$}{
		Obtain the optimal $\alpha_m$ to (P2') with given $\{\mv{Q}[n]\}_{n=1}^N$ and $\{\alpha_i,i\neq m\}_{i=1}^M$ via CVX.
	}
	Obtain the optimal $\{{\mv{Q}}[n]\}_{n=1}^N$ to (P2') with given $\{{\alpha}_m\}_{m=1}^M$ via CVX.\\
	Check convergence. If yes, stop; if not, go to Step 4.\\
	Set ${\mv{\phi}}=\mathrm{diag}\{\alpha_1/|\alpha_1|,...,\alpha_M/|\alpha_M|\}$. \\
	Obtain the optimal $\{{\mv{Q}}[n]\}_{n=1}^N$ to (P2) with given $\mv{\phi}$ via CVX.
\end{algorithm}
	\begin{figure}[b]
	\vspace{-8mm}
		\centering
	\includegraphics[width=6cm]{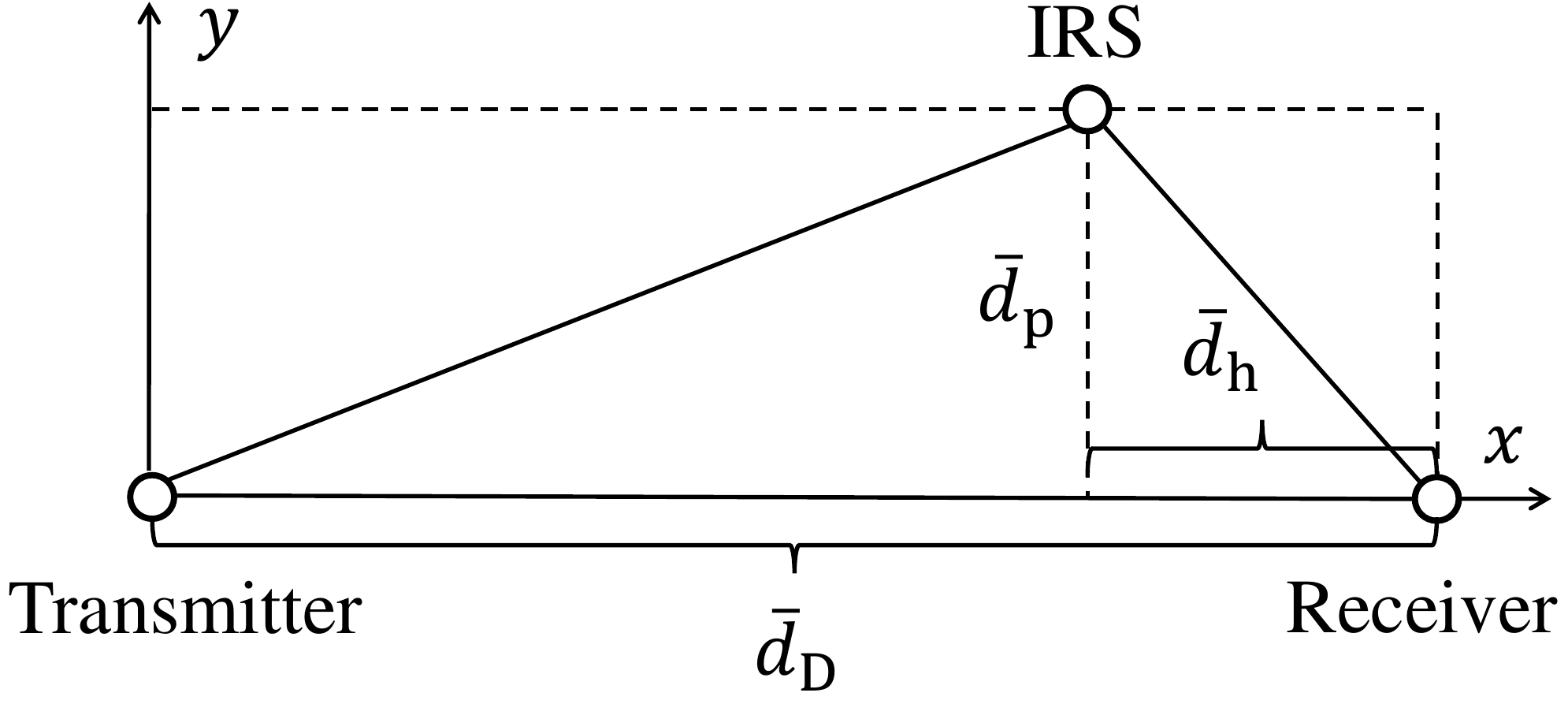}
	\vspace{-3mm}
	\caption{Horizontal locations of the transmitter, IRS, and receiver.}\label{Setup}
\end{figure}
\begin{table}[t]
		\centering
	\caption{Summary of System Setup and Simulation Parameters}\label{table_parameter}
	\vspace{-3mm}
	\resizebox{\textwidth}{!}{\begin{tabular}{|c|c|c|c|}
			\hline 
			& Direct link, $\mv{H}$ &  Transmitter-IRS link, $\mv{T}$ & IRS-Receiver link, $\mv{R}$\\ 
			\hline
			Distance (m) & $d_{\mathrm{D}}=\sqrt{\bar{d}_{\mathrm{D}}^2+\bar{H}^2}$ & $d_{\mathrm{TI}}=\sqrt{(\bar{d}_{\mathrm{D}}-\bar{d}_{\mathrm{h}})^2+\bar{d}_{\mathrm{p}}^2}$ & $d_{\mathrm{IR}}=\sqrt{\bar{d}_{\mathrm{h}}^2+\bar{d}_{\mathrm{p}}^2+\bar{H}^2}$ \\
			\hline
			AoA & $\theta_{\mathrm{D}}^{\mathrm{A}}=0$ & $\theta_{\mathrm{TI}}^{\mathrm{A}}=\arctan\left(\frac{\bar{d}_{\mathrm{D}}-\bar{d}_{\mathrm{h}}}{\bar{d}_{\mathrm{p}}}\right)$, $\psi_{\mathrm{TI}}^{\mathrm{A}}=0$ & $\theta_{\mathrm{IR}}^{\mathrm{A}}=\arctan\left(\frac{\bar{d}_{\mathrm{h}}}{\bar{d}_{\mathrm{p}}}\right)$ \\
			\hline 
			AoD & $\theta_{\mathrm{D}}^{\mathrm{D}}=0$ & $\theta_{\mathrm{TI}}^{\mathrm{D}}=\frac{\pi}{2}-\theta_{\mathrm{TI}}^{\mathrm{A}}$ &  $\theta_{\mathrm{IR}}^{\mathrm{D}}=\frac{\pi}{2}-\theta_{\mathrm{IR}}^{\mathrm{A}}$, $\psi_{\mathrm{IR}}^{\mathrm{D}}=\arctan\left(\frac{-\bar{H}}{\sqrt{\bar{d}_{\mathrm{p}}^2+\bar{d}_{\mathrm{h}}^2}}\right)$  \\
			\hline
			LoS component & $\mv{H}_{\mathrm{LoS}}=\mv{\mathrm{a}}_{\mathrm{R}}(\theta_{\mathrm{D}}^{\mathrm{A}})\mv{\mathrm{a}}_{\mathrm{T}}(\theta_{\mathrm{D}}^{\mathrm{D}})^H$ & $\mv{T}_{\mathrm{LoS}}=\mv{\mathrm{a}}_{\mathrm{I}}(\theta_{\mathrm{TI}}^{\mathrm{A}},\psi_{\mathrm{TI}}^{\mathrm{A}})\mv{\mathrm{a}}_{\mathrm{T}}(\theta_{\mathrm{TI}}^{\mathrm{D}})^H$ &  $\mv{R}_{\mathrm{LoS}}=\mv{\mathrm{a}}_{\mathrm{R}}(\theta_{\mathrm{TI}}^{\mathrm{A}})\mv{\mathrm{a}}_{\mathrm{I}}(\theta_{\mathrm{TI}}^{\mathrm{D}},\psi_{\mathrm{TI}}^{\mathrm{D}})^H$  \\
			\hline
	\end{tabular}} 
	\vspace{-9mm}
\end{table}
\vspace{-8mm}
\section{Numerical Results}\label{sec_num}
\vspace{-1mm}
In this section, we provide numerical results to examine the performance of our proposed algorithms for maximizing the IRS-aided MIMO system capacities. Under a three-dimensional (3D) Cartesian coordinate system, we assume that both the transmitter and the receiver are equipped with a uniform linear array (ULA) located on the $y$-axis with antenna spacing $d_{\mathrm{A}}=\lambda/2$, where $\lambda$ denotes the wavelength; while the IRS is equipped with a uniform planar array (UPA) located parallel to the $x-z$ plane with IRS element spacing $d_{\mathrm{I}}=\lambda/8$. For illustration, we consider a scenario where the transmitter and the receiver serve as a base station (BS) and a cell-edge user, respectively, and assume that both the BS and the IRS are located at the same altitude above the user by $\bar{H}$ meter (m). The locations of the reference antenna/element at the transmitter, the IRS, and the receiver are set as $(0,0,\bar{H})$, $(\bar{d}_{\mathrm{D}}-\bar{d}_{\mathrm{h}},\bar{d}_{\mathrm{p}},\bar{H})$, and $(\bar{d}_{\mathrm{D}},0,0)$, respectively, whose horizontal projections are illustrated in Fig. \ref{Setup}. The 3D distances for the direct link, the transmitter-IRS link, and the IRS-receiver link can be thus obtained as $d_{\mathrm{D}}$, $d_{\mathrm{TI}}$, and $d_{\mathrm{IR}}$ given in Table \ref{table_parameter}, respectively. We further set $\bar{H}=10$ m, and the distances from the projection of the IRS on the $x$-axis to the IRS and the receiver as $\bar{d}_{\mathrm{p}}=2$ m and $\bar{d}_{\mathrm{h}}=2$ m, respectively, since the IRS is practically deployed in the user's close vicinity to improve its performance. The distance-dependent path loss for all channels is modeled as $\beta=\beta_0(d/d_0)^{-\bar{\alpha}}$, where $\beta_0=-30$ dB denotes the path loss at the reference distance $d_0=1$ m; $\bar{\alpha}$ denotes the path loss exponent. Under this model, we denote $\beta_{\mathrm{D}}$, $\beta_{\mathrm{TI}}$, and $\beta_{\mathrm{IR}}$ as the path loss of the direct link, the transmitter-IRS link, and the IRS-receiver link, respectively; the path loss exponents for the corresponding links are set as $\bar{\alpha}_{\mathrm{D}}=3.5$, $\bar{\alpha}_{\mathrm{TI}}=2.2$, and $\bar{\alpha}_{\mathrm{IR}}=2.8$, respectively. The specific channel models for the frequency-flat and frequency-selective cases will be given in Section \ref{sec_num_flat} and Section \ref{sec_num_selective}, respectively. We consider a noise power spectrum density of $-169$ dBm/Hz with additional $9$ dB noise figure, and a system bandwidth of $10$ MHz, which yields $\sigma^2=-90$ dBm for narrowband MIMO systems and $\bar{\sigma}^2=-90-10\log_{10}N_f$ dBm for MIMO-OFDM systems with $N_f$ subcarriers. Unless specified otherwise, we set the transmit power constraint as $P=30$ dBm for the considered downlink transmission. For all the proposed alternating optimization algorithms, we set the number of random initializations as $L=100$, and the convergence threshold in terms of the relative increment in the objective value as $\epsilon=10^{-5}$. All the results are averaged over $100$ independent channel realizations.
\vspace{-4mm}
\subsection{MIMO System under Frequency-Flat Channel}\label{sec_num_flat}
\vspace{-1mm}
To start with, we consider narrowband MIMO systems under frequency-flat channels. We adopt the \emph{Rician fading} model for all channels, i.e., $\mv{H}$, ${\mv{T}}$, and ${\mv{R}}$. The direct channel $\mv{H}$ is modeled as ${\mv{H}}=\sqrt{\beta_{\mathrm{D}}/(K_{\mathrm{D}}+1)}\left(\sqrt{K_{\mathrm{D}}}\mv{H}_{\mathrm{LoS}}+\mv{H}_{\mathrm{NLoS}}\right)$, where $\mv{H}_{\mathrm{LoS}}$ denotes the LoS component, as will be specified below; $\mv{H}_{\mathrm{NLoS}}$ denotes the NLoS component modeled by Rayleigh fading, with $[\mv{H}]_{i,j}\sim \mathcal{CN}(0,1),\forall i,j$; and $K_{\mathrm{D}}\in [0,\infty)$ denotes the Rician factor. Note that by considering different $K_{\mathrm{D}}$, this model corresponds to various practical channels including the deterministic LoS channel when $K_{\mathrm{D}}\rightarrow\infty$, and the Rayleigh fading channel when $K_{\mathrm{D}}=0$. The transmitter-IRS channel ${\mv{T}}$ and the IRS-receiver channel ${\mv{R}}$ are similarly modeled with Rician factors $K_{\mathrm{TI}}$ and $K_{\mathrm{IR}}$, respectively. Specifically, the LoS component for each channel is modeled as the product of the array responses at two sides. For the ULA at the transmitter, the array response is modeled as ${\mv{\mathrm{a}}}_{\mathrm{T}}(\theta)\in \mathbb{C}^{N_t\times 1}$, with $[\mv{\mathrm{a}}_{\mathrm{T}}(\theta)]_{n}=e^{j2\pi(n-1)d_{\mathrm{A}}\sin \theta/\lambda},\forall n$, where $\theta\in [0,2\pi)$ denotes the angle-of-arrival (AoA) or angle-of-departure (AoD) \cite{Phased}; the array response for the ULA at the receiver with AoA/AoD $\theta$ is similarly modeled as ${\mv{\mathrm{a}}}_{\mathrm{R}}(\theta)\in \mathbb{C}^{N_r\times 1}$, with $[\mv{\mathrm{a}}_{\mathrm{R}}(\theta)]_{n}=e^{j2\pi(n-1)d_{\mathrm{A}}\sin \theta/\lambda},\forall n$. For the UPA at the IRS, the array response is modeled as $\mv{\mathrm{a}}_{\mathrm{I}}(\theta,\psi)\in \mathbb{C}^{M\times 1}$ with $[\mv{\mathrm{a}}_{\mathrm{I}}(\theta,\psi)]_m=e^{j2\pi d_{\mathrm{I}}(\lfloor \frac{m}{M_x} \rfloor\sin \psi\sin \theta+(m-\lfloor \frac{m}{M_x} \rfloor M_x)\sin\psi\cos\theta)/\lambda},\forall m$, where $\theta\in [0,2\pi)$ and $\psi\in [-\pi/2,\pi/2)$ denote the azimulth AoA/AoD and elevation AoA/AoD, respectively \cite{Phased}; $M_x$ denotes the number of IRS elements in each row along the $x$-axis set as $M_x=\min(M,10)$ in the sequel; and $\lfloor x\rfloor$ denotes the maximum integer no larger than a real number $x$. In Table \ref{table_parameter}, we summarize the AoA/AoD for all channels involved under the considered setup and their corresponding LoS components. In the following, we consider a MIMO system with $N_t=N_r=4$ for evaluating the performance of our proposed Algorithm \ref{algo1}.
\begin{figure}[t]
	\centering
	\includegraphics[width=8cm]{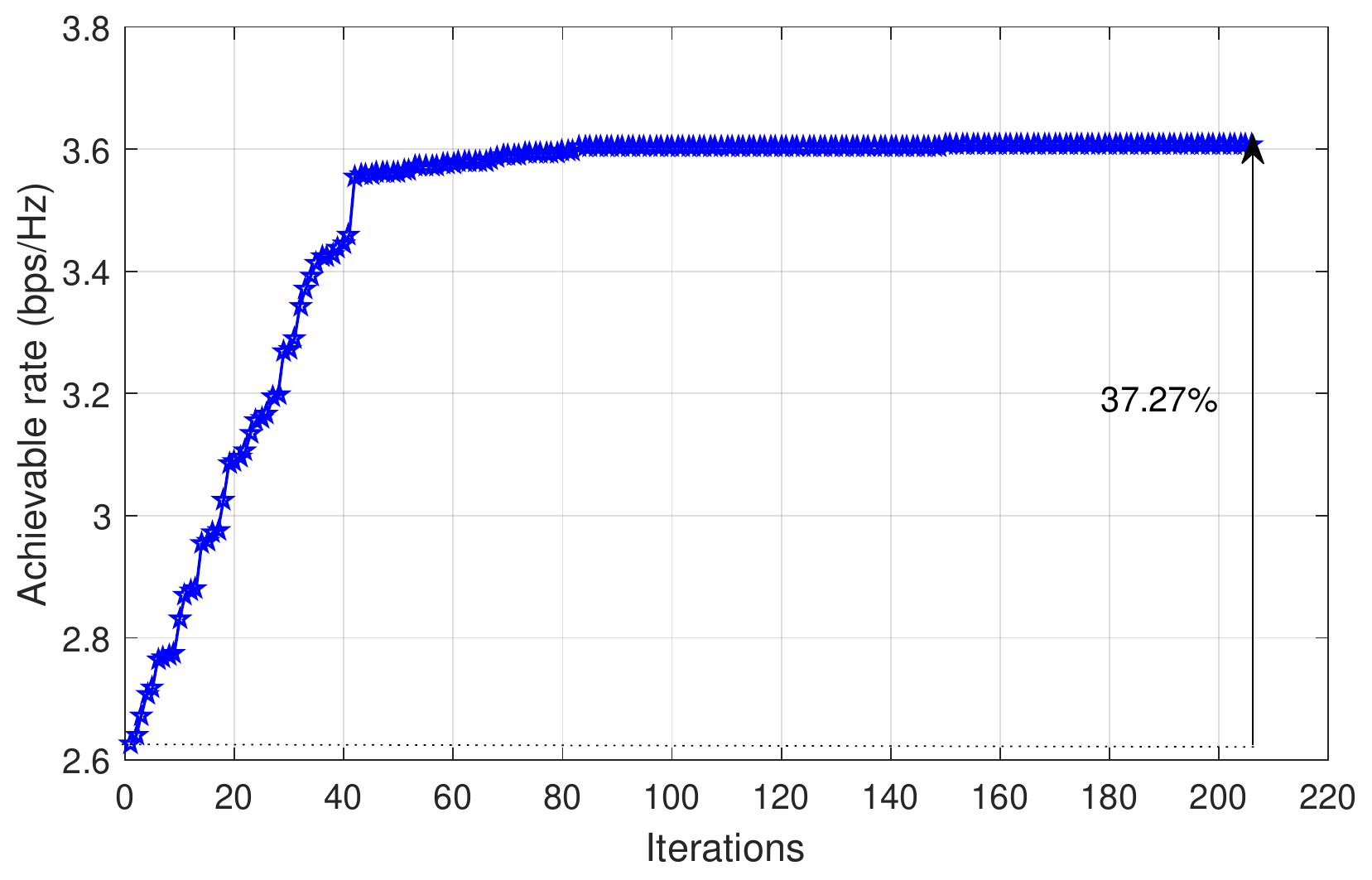}
	\vspace{-4mm}	
	\caption{Convergence behavior of Algorithm \ref{algo1} with $M=40$.}\label{MIMO_Convergence}
	\vspace{-9mm}
\end{figure}

First, we set $M=40$, $\bar{d}_{\mathrm{D}}=600$ m, $K_{\mathrm{D}}=K_{\mathrm{TI}}=K_{\mathrm{IR}}=0$, and show in Fig. \ref{MIMO_Convergence} the convergence behavior of Algorithm \ref{algo1} (under one channel realization). It is observed that Algorithm \ref{algo1} converges monotonically, which validates our analysis in Section \ref{sec_solution}, and the convergence speed is fast (about $5$ outer iterations to be within the considered high precision). Moreover, the converged rate is increased by $37.27\%$ as compared to that at the initial point. Next, under $K_{\mathrm{D}}=K_{\mathrm{TI}}=K_{\mathrm{IR}}=0$, we compare the performance of Algorithm \ref{algo1} with the following benchmark schemes:
\begin{enumerate}
	\item {\bf{Without IRS}}: Obtain the channel capacity in (\ref{capacity}) by optimizing $\mv{Q}$ with given $\tilde{\mv{H}}={\mv{H}}$.
	\item {\bf{Random phase}}: Randomly generate $\{\alpha_m\}_{m=1}^M$ with $|\alpha_m|=1,\forall m$ and phases of $\alpha_m$'s following independent uniform distribution in $[0,2\pi)$. Obtain the channel capacity in (\ref{capacity}) by optimizing $\mv{Q}$ with given $\{\alpha_m\}_{m=1}^M$.
	\item {\bf{Strongest eigenchannel power maximization}}: The alternative algorithm presented in Section \ref{sec_alternativelow} customized for the low-SNR regime.
	\item {\bf{Channel total power maximization}}: The alternative algorithm presented in Section \ref{sec_alternativehigh} customized for the high-SNR regime.
	\item {\bf{Reflection optimization with fixed $\mv{Q}$}}: In this scheme, we first obtain the optimal (capacity-achieving) transmit covariance matrix $\mv{Q}$ for the direct channel $\mv{H}$; then, we apply Algorithm \ref{algo1} to optimize $\{\alpha_m\}_{m=1}^M$ with $\mv{Q}$ fixed. 
	\item {\bf{Heuristic channel total power maximization}}: In this scheme, we propose a heuristic approach to maximize the channel total power by maximizing its lower bound, which is given by $\|\tilde{\mv{H}}_F\|^2=\sum_{i=1}^{N_r}\sum_{j=1}^{N_t}|[\mv{H}]_{i,j}+\sum_{m=1}^M \alpha_mr_{mi}t_{mj}^*|^2\geq |\tilde{h}^d+\sum_{m=1}^M \alpha_m \tilde{h}^r_m|^2$, where $\tilde{h}^d\overset{\Delta}{=}\sum_{i=1}^{N_r}\sum_{j=1}^{N_t}[\mv{H}]_{i,j}$ and $\tilde{h}^r_m\overset{\Delta}{=}\sum_{i=1}^{N_r}\sum_{j=1}^{N_t}r_{mi}t_{mj}^*$. The optimal $\{\alpha_m\}_{m=1}^M$ that maximizes this lower bound can be easily shown to be $\alpha_m=e^{j(\arg\{\tilde{h}^d\}-\arg\{\tilde{h}^r_m\})},\forall m$, which can be obtained with complexity $\mathcal{O}(N_rN_tM)$. Note that this scheme can be easily shown to be the optimal solution to the SISO case with $N_t=N_r=1$.
\end{enumerate}
\begin{figure}[b]
	\vspace{-12mm}
	\centering
	\subfigure[Achievable rate versus $M$]{
		\includegraphics[height=5.5cm]{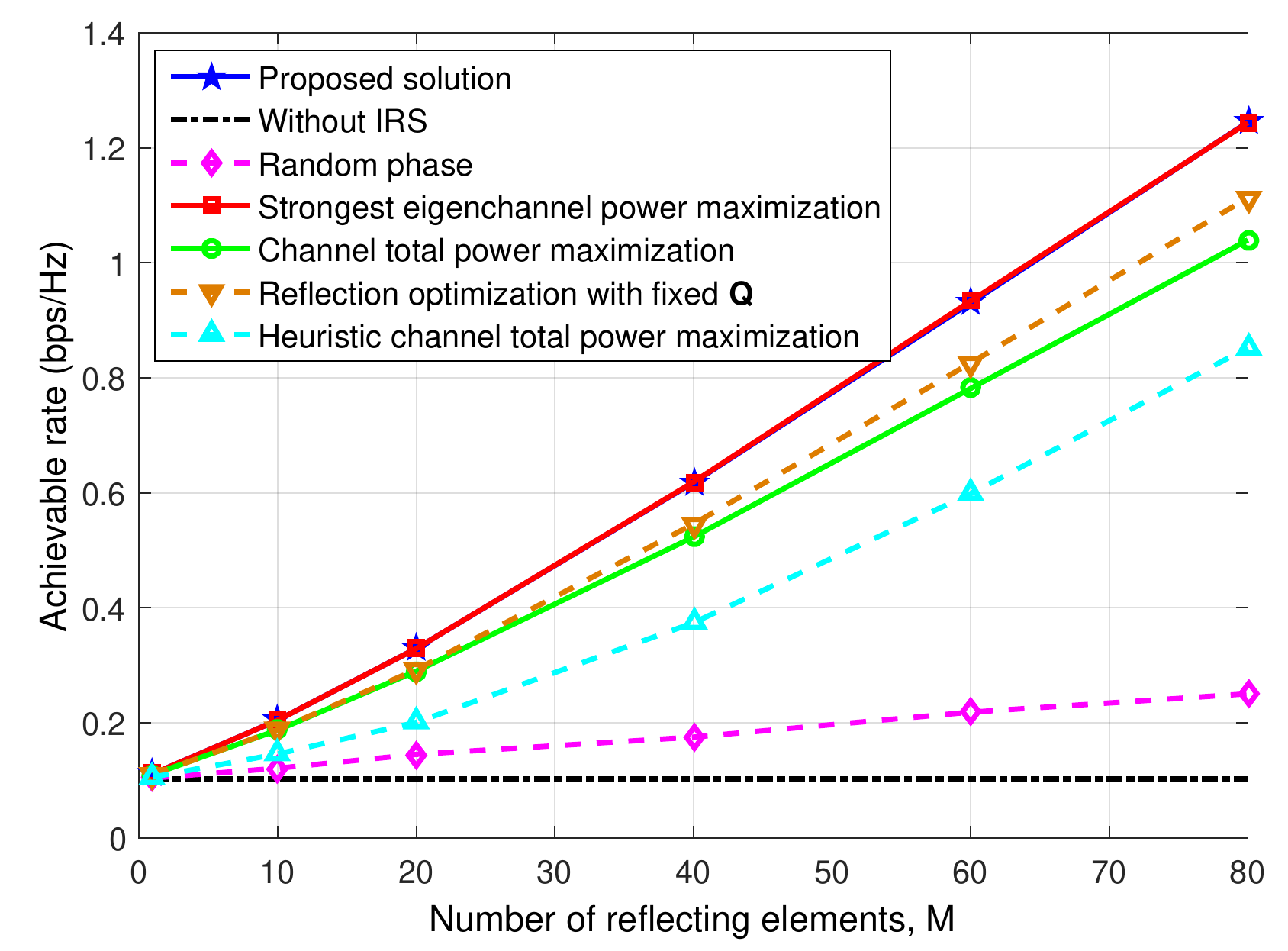}}
	\subfigure[Strongest eigenchannel power versus $M$]{
		\includegraphics[height=5.5cm]{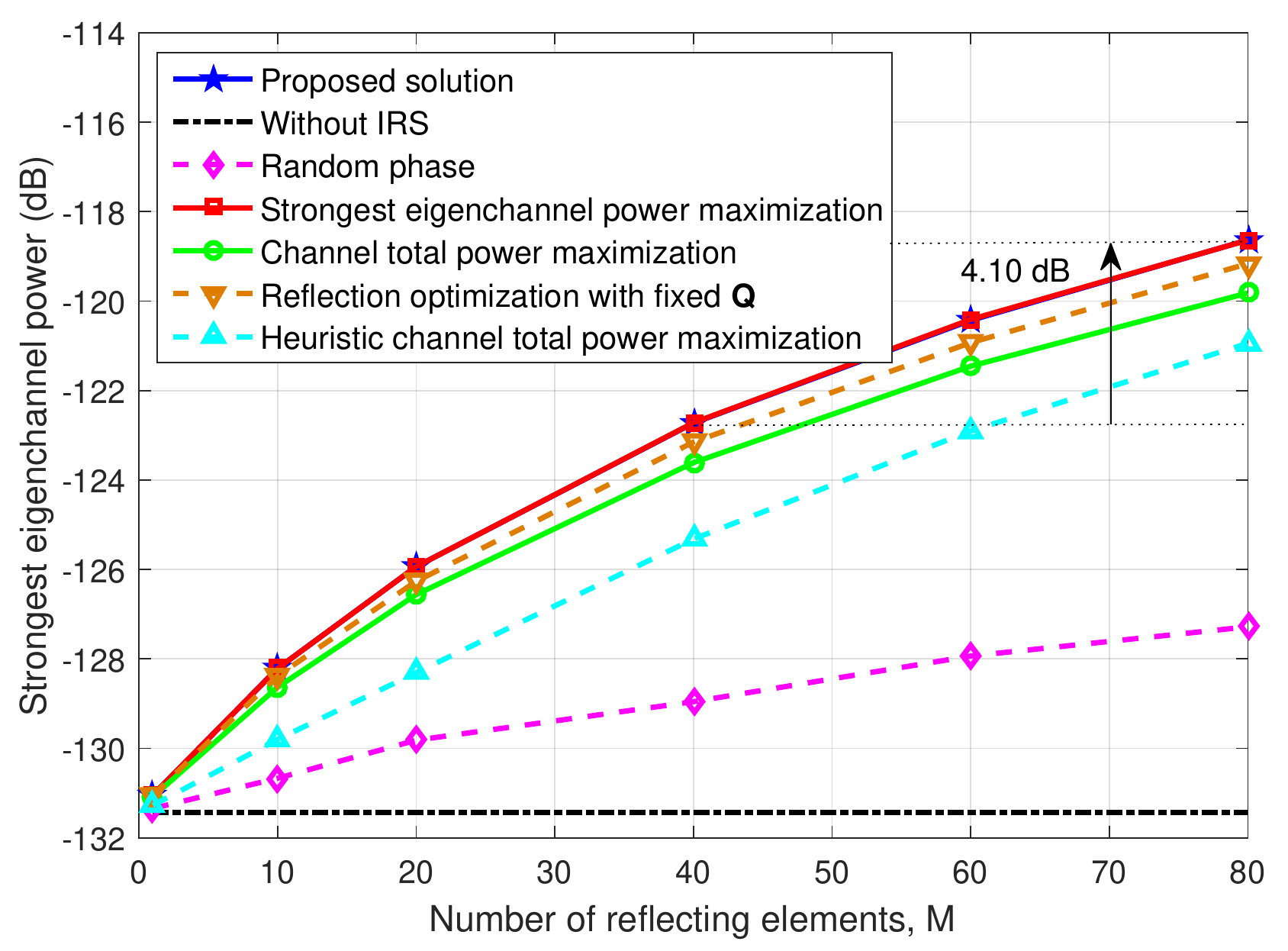}}
	\vspace{-3mm}
	\caption{Performance of IRS-aided MIMO communication in the low-SNR regime ($\bar{d}_{\mathrm{D}}=1500$ m).}\label{Fig_LowSNR}
\end{figure}

\begin{figure}[t]
	\centering
	\subfigure[Achievable rate versus $M$]{
		\includegraphics[height=4cm]{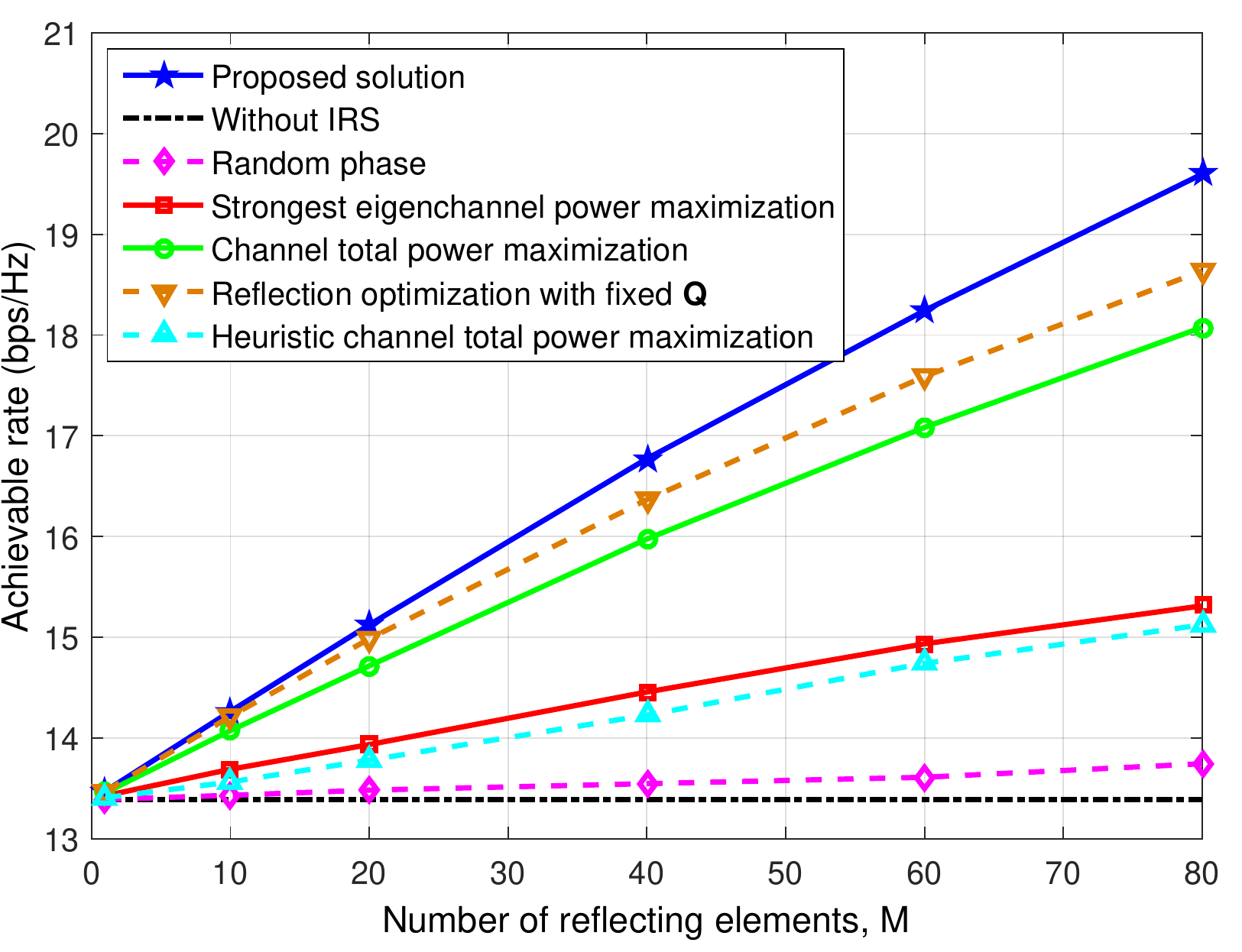}}
	\hspace{-3mm}
	\subfigure[Channel total power versus $M$]{
		\includegraphics[height=4cm]{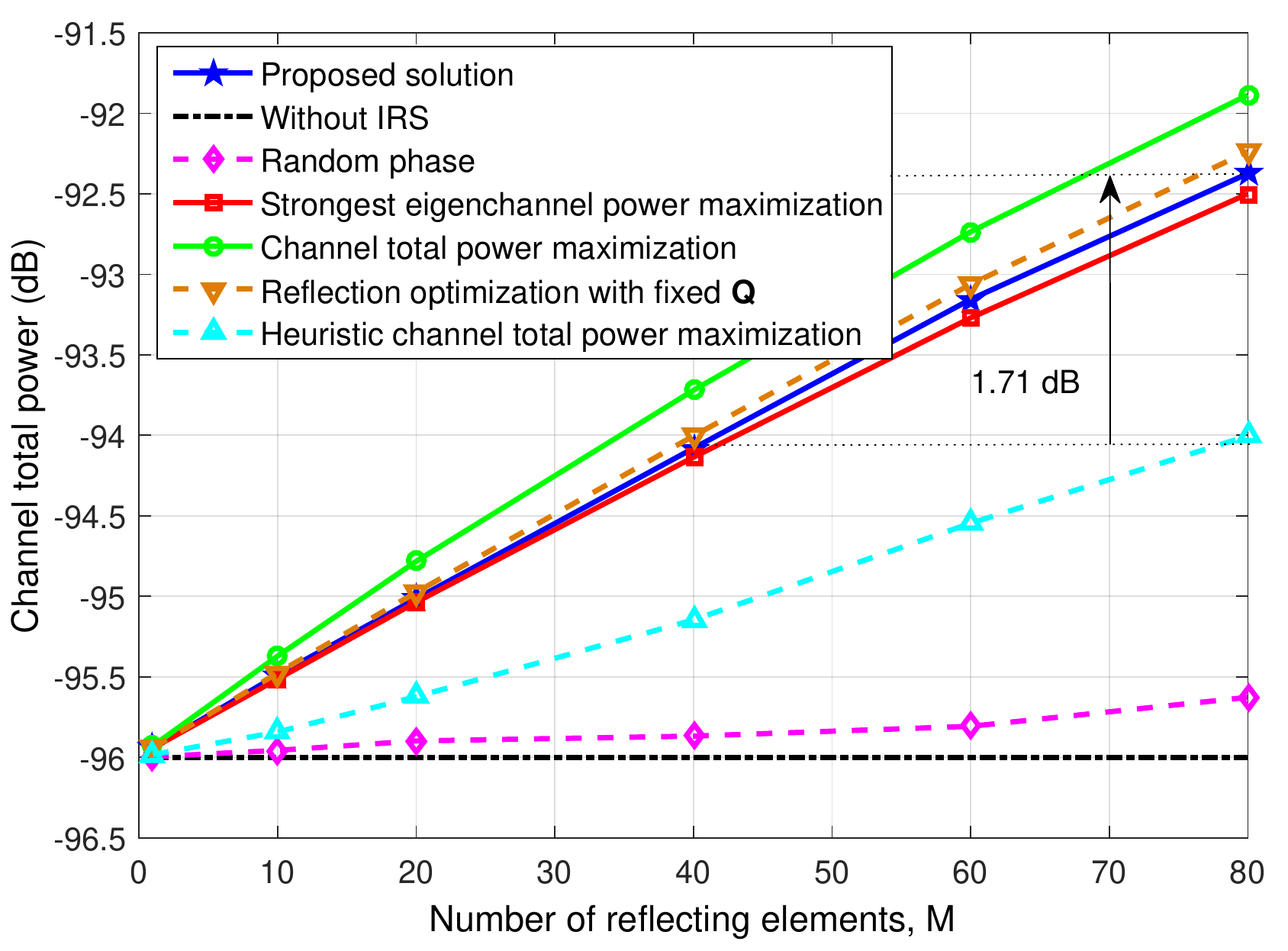}}
	\hspace{-3mm}
	\subfigure[Channel condition number versus $M$]{
		\includegraphics[height=4cm]{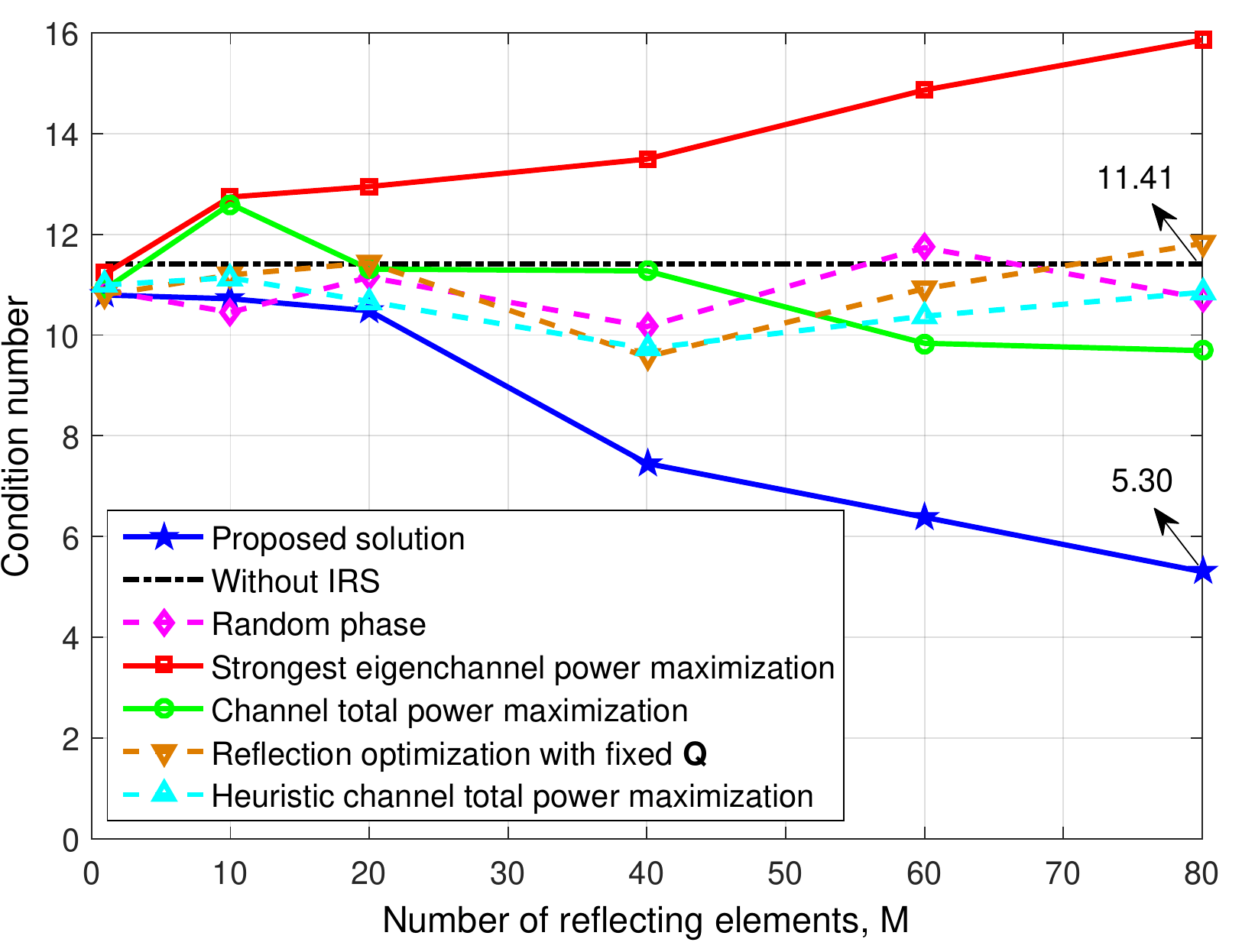}}
	\vspace{-3mm}
	\caption{Performance of IRS-aided MIMO communication in the high-SNR regime ($\bar{d}_{\mathrm{D}}=170$ m).}\label{Fig_HighSNR}
	\vspace{-10mm}
\end{figure}

For illustration, we consider two transmitter-receiver horizontal distances given by $\bar{d}_{\mathrm{D}}=1500$ m and $\bar{d}_{\mathrm{D}}=170$ m, which correspond to the \emph{low-SNR regime} and \emph{high-SNR regime}, respectively. For these two cases, we show in Fig. \ref{Fig_LowSNR} (a) and Fig. \ref{Fig_HighSNR} (a) respectively the achievable rate versus the number of reflecting elements $M$ for the proposed algorithm (Algorithm \ref{algo1}) and the benchmark schemes, respectively. For both SNR regimes, it is observed that all the schemes with IRS outperform that without the IRS, and the performance gain increases with $M$; moreover, schemes 3)-6) with IRS all outperform the random phase scheme. It is also observed that our proposed algorithm achieves the best performance among all schemes in both SNR regimes and at all values of $M$. Particularly, the proposed algorithm outperforms benchmark scheme 5) with transmit covariance matrix $\mv{Q}$ optimized only based on the direct MIMO channel, which shows the necessity of jointly optimizing $\mv{Q}$ and the IRS reflection coefficients.

Moreover, in the low-SNR regime, it is observed from Fig. \ref{Fig_LowSNR} (a) that the strongest eigenchannel power maximization algorithm achieves almost the same performance as the proposed algorithm, which is consistent with our results in Section \ref{sec_alternativelow}; thus, this scheme can serve as a suitable low-complexity alternative to our proposed Algorithm \ref{algo1} in the low-SNR regime. To draw more insight, we further depict in Fig. \ref{Fig_LowSNR} (b) the (average) strongest eigenchannel power $[\tilde{\mv{\Lambda}}]_{\max}^2$ for each scheme over $M$. It is observed that the increment of $[\tilde{\mv{\Lambda}}]_{\max}^2$ for all schemes with IRS over that without IRS increases as $M$ increases, while our proposed algorithm and the strongest eigenchannel power maximization algorithm are able to boost $[\tilde{\mv{\Lambda}}]_{\max}^2$ by $4.10$ dB by doubling $M$ from $40$ to $80$. On the other hand, in the high-SNR regime, it is observed from Fig. \ref{Fig_HighSNR} (a) that the channel total power maximization algorithm achieves close performance to the proposed Algorithm \ref{algo1}, which validates the effectiveness of adopting the channel total power as an approximate performance metric in the high-SNR regime. However, a performance gap between the two schemes still exists, which is further investigated as follows. Note from (\ref{capacity_highSNR}) that besides the channel total power, there are two other key channel parameters that also influence the capacity in the high-SNR regime, namely, the channel \emph{rank} $D$, and the channel \emph{condition number} $\kappa=\frac{[\tilde{\mv{\Lambda}}]_{\max}}{[\tilde{\mv{\Lambda}}]_{\min}}$, with $[\tilde{\mv{\Lambda}}]_{\min}$ denoting the minimum singular value of $\tilde{\mv{H}}$, where the capacity generally increases with $D$ and decreases with $\kappa$ \cite{Fundamental}. Motiaved by this, we show in Fig. \ref{Fig_HighSNR} (b) and (c) the (average) channel total power $\|\tilde{\mv{H}}\|_F^2$ and condition number $\kappa$ of the effective channels resulted from the various schemes, where the rank of the effective channel is observed to be always full (i.e., $D=\min(N_t,N_r)$) for each scheme. It is further observed from Fig. \ref{Fig_HighSNR} (b) that the channel total power of our proposed scheme 4) increases most significantly with $M$. On the other hand, it is observed from Fig. \ref{Fig_HighSNR} (c) that the proposed Algorithm \ref{algo1} generally yields the smallest condition number among all schemes, which decreases as $M$ increases. The low condition number yields more balanced power distribution over the eigenchannels, thus leading to its performance gain over the channel total power maximization scheme 4) despite that the latter achieves larger channel total power shown in Fig. \ref{Fig_HighSNR} (b). The above results indicate the our proposed Algorithm \ref{algo1} is able to reshape the MIMO channel towards more favorable condition for capacity maximization, by striking a balance between maximizing the channel total power and minimizing the channel condition number.
\begin{figure}[t]
	\centering
	\subfigure[$K_{\mathrm{TI}}\rightarrow\infty$]{
		\includegraphics[height=5.5cm]{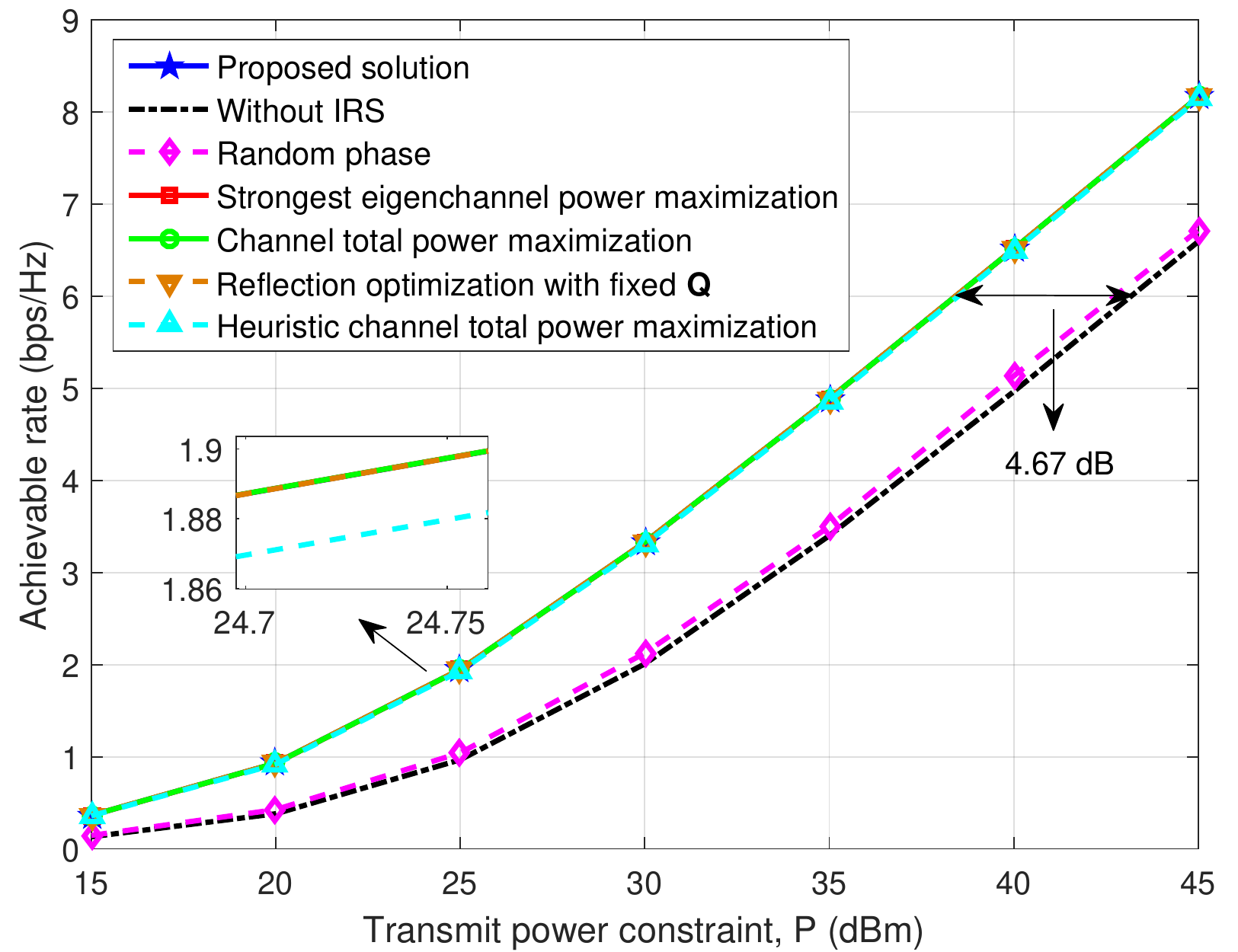}}
	\subfigure[$K_{\mathrm{TI}}=1$]{
		\includegraphics[height=5.5cm]{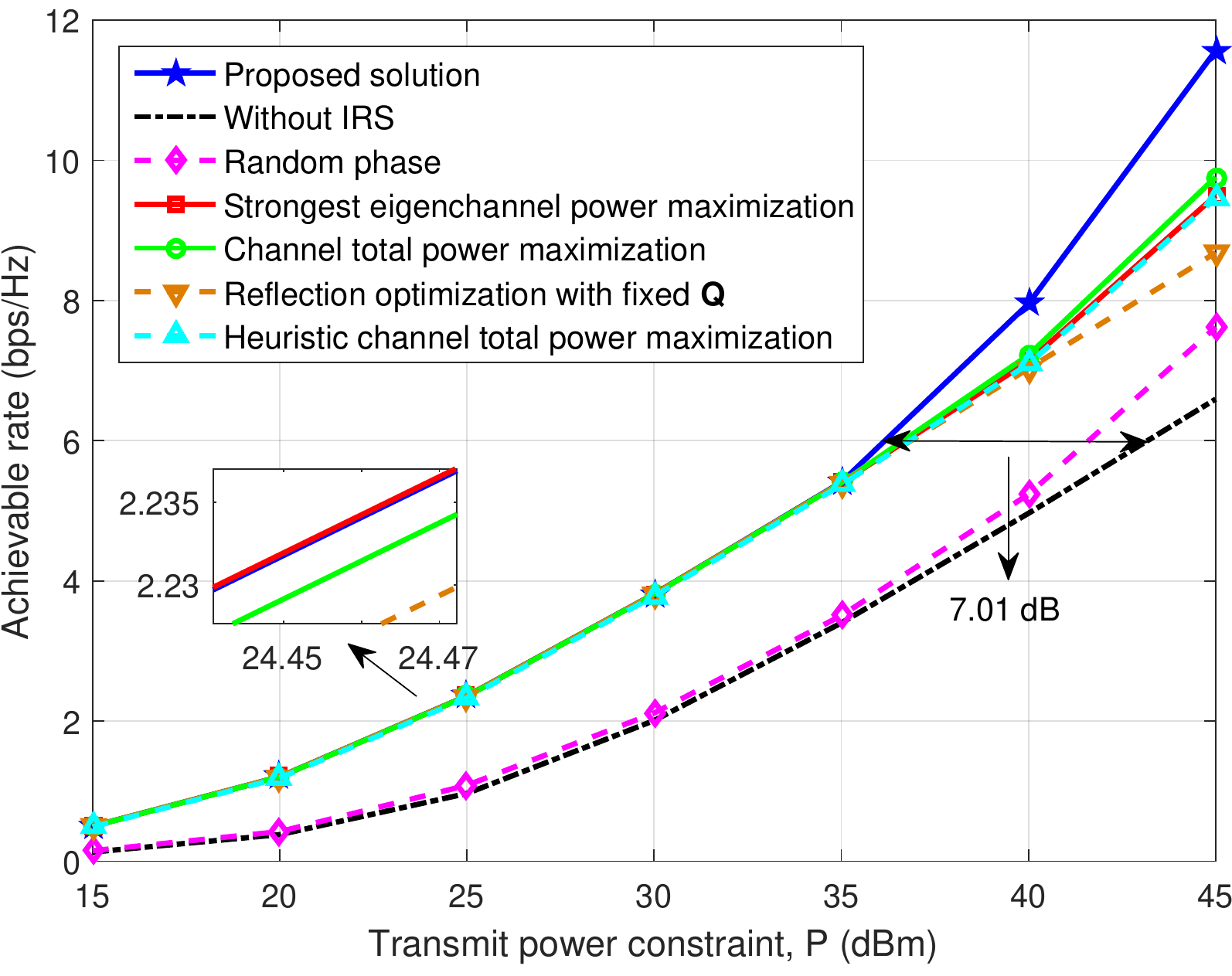}}
	\vspace{-3mm}
	\caption{Achievable rate versus transmit power constraint $P$ for IRS-aided MIMO communication with $M=40$.}\label{MIMO_SNR}
	\vspace{-10mm}
\end{figure}

Last, we consider a setup with $M=40$ and $\bar{d}_{\mathrm{D}}=600$ m, an LoS direct channel with $K_{\mathrm{D}}\rightarrow \infty$, and two types of transmitter-IRS channels with $K_{\mathrm{TI}}\rightarrow\infty$ (i.e., LoS) and $K_{\mathrm{TI}}=1$, respectively. In Fig. \ref{MIMO_SNR}, we show the achievable rate of the various schemes versus the transmit power constraint $P$. It is observed that to achieve a given rate, our proposed algorithm requires dramatically reduced transmit power as compared to the scheme without IRS. Moreover, it is observed from Fig. \ref{MIMO_SNR} that with $K_{\mathrm{TI}}\rightarrow\infty$, the proposed algorithm achieves similar performance with various lower-complexity benchmark schemes, and all the schemes yield a similar spatial multiplexing gain as $P$ increases. This is because with $K_{\mathrm{D}}\rightarrow \infty$ and $K_{\mathrm{TI}}\rightarrow\infty$, both the direct channel and the transmitter-IRS channel are of rank one, which are highly correlated since the IRS is placed in the vicinity of the receiver. This makes the overall effective MIMO channel rank-one regardless of the reflection coefficient design, which thus can only support one data stream. On the other hand, it is observed that for the case of $K_{\mathrm{TI}}=1$, the proposed algorithm as well as other benchmark schemes yield a larger spatial multiplexing gain compared to the scheme without IRS (i.e., $4$ for the proposed algorithm), since the NLoS component in $\mv{T}$ can be leveraged to create a higher-rank effective MIMO channel via proper reflection coefficient design. This shows that IRS can be used to effectively improve the MIMO channel rank and hence the spatial multiplexing gain by deploying it in a rich-scattering environment.
\vspace{-5mm}
\begin{figure}[b]
	\vspace{-10mm}
	\centering
	\subfigure[$N=8$]{
		\includegraphics[height=5.5cm]{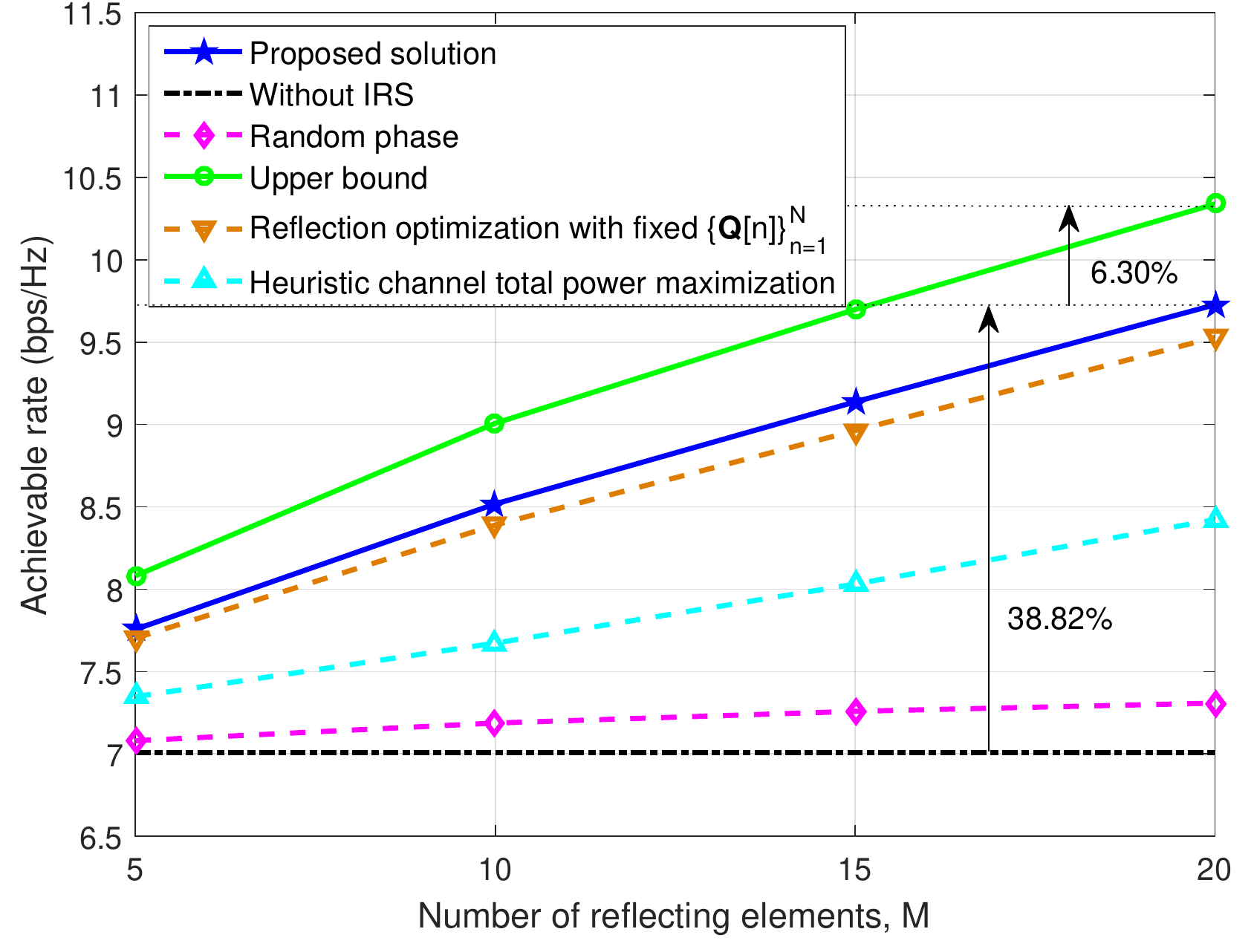}}
	\subfigure[$N=32$]{
		\includegraphics[height=5.5cm]{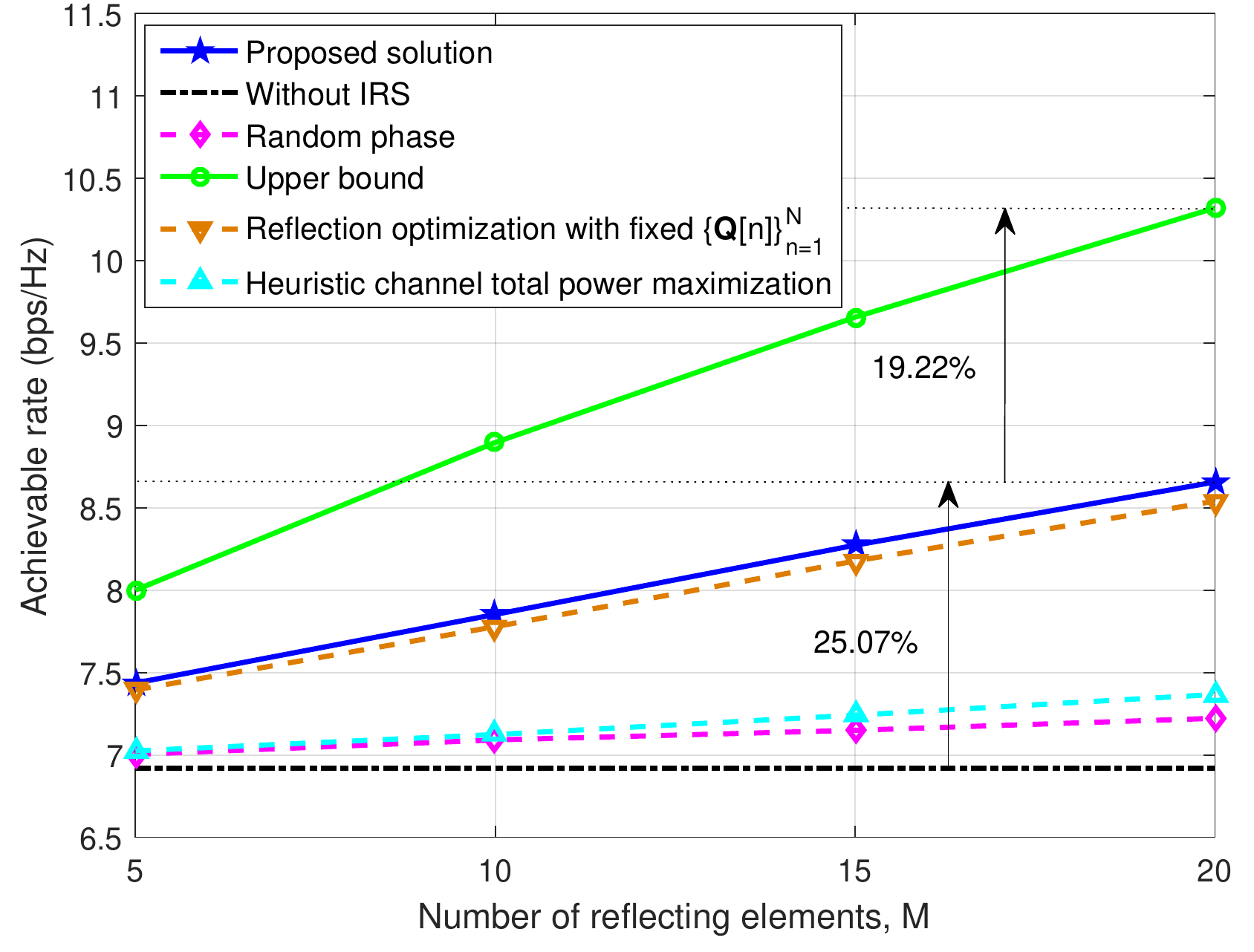}}
	\vspace{-3mm}
	\caption{Achievable rate versus $M$ for IRS-aided MIMO-OFDM communication.}\label{Fig_OFDM}
\end{figure}
\subsection{MIMO-OFDM System under Frequency-Selective Channel}\label{sec_num_selective}
\vspace{-1mm}
Next, we evaluate the performance of Algorithm \ref{algo_OFDM} for MIMO-OFDM systems with $N_t=N_r=2$, $\bar{d}_{\mathrm{D}}=800$ m, $N_f=512$ and $\mu=128$. We consider two sets of parameters, namely, $N=8$, $L_{\mathrm{D}}=2$, $L_{\mathrm{TI}}=1$, $L_{\mathrm{IR}}=1$, and $N=32$, $L_{\mathrm{D}}=8$, $L_{\mathrm{TI}}=4$, $L_{\mathrm{IR}}=4$. The time-domain channels $\{\bar{\mv{H}}_l\}_{l=0}^{L_{\mathrm{D}}-1}$, $\{\bar{\mv{T}}_l\}_{l=0}^{L_{\mathrm{TI}}-1}$ and $\{\bar{\mv{R}}_l\}_{l=0}^{L_{\mathrm{IR}}-1}$ are assumed to be independent random matrices, each consisting of entries independently distributed as $\mathcal{CN}(0,\beta_{\mathrm{D}}/{L}_{\mathrm{D}})$, $\mathcal{CN}(0,\beta_{\mathrm{TI}}/{L}_{\mathrm{TI}})$ and $\mathcal{CN}(0,\beta_{\mathrm{IR}}/{L}_{\mathrm{IR}})$, respectively. For comparison, we consider a performance upper bound where the reflecting elements at the IRS are assumed to be able to be adjusted for different subcarriers, which is however difficult to implement in practice. In this case, $N$ sets of reflection coefficients $\{\alpha_m[n],m\in \mathcal{M}\}_{n=1}^N$ can be designed in parallel based on Algorithm \ref{algo1} as in the narrowband case. Moreover, similar to benchmark scheme 5) in the narrowband case, we consider a benchmark scheme where only the reflection coefficients are optimized via Algorithm \ref{algo_OFDM} with the transmit covariance matrices $\{\mv{Q}[n]\}_{n=1}^N$ fixed as the optimal solution for the direct channel. Furthermore, motivated by benchmark scheme 6) in the narrowband case, we consider another benchmark scheme to heuristically maximize the MIMO-OFDM channel total power, which is given by 
$\sum_{n=1}^N\|\tilde{\mv{H}}[n]\|_F^2=\sum_{n=1}^N\sum_{i=1}^{N_r}\sum_{j=1}^{N_t}|[\mv{H}[n]]_{i,j}+\sum_{m=1}^M \alpha_mr_{mi}[n]t_{mj}[n]^*|^2\geq |\tilde{h}^d+\sum_{m=1}^M \alpha_m \tilde{h}^r_m|^2$, where $\tilde{h}^d\overset{\Delta}{=}\sum_{n=1}^N\sum_{i=1}^{N_r}\sum_{j=1}^{N_t}[\mv{H}[n]]_{i,j}$ and $\tilde{h}^r_m\overset{\Delta}{=}\sum_{n=1}^N\sum_{i=1}^{N_r}\sum_{j=1}^{N_t}r_{mi}[n]t_{mj}[n]^*$. The optimal $\{\alpha_m\}_{m=1}^M$ that maximizes this lower bound can be shown to be $\alpha_m=e^{j(\arg\{\tilde{h}^d\}-\arg\{\tilde{h}^r_m\})},\forall m$.
In addition, we also consider the scheme without IRS and that with random phase as two other benchmark schemes.

In Fig. \ref{Fig_OFDM}, we show the achievable rate of the proposed and benchmark schemes versus $M$ for the case of $N=8$ and $N=32$, respectively. It is observed that our proposed algorithm is able to achieve superior achievable rate compared to the case without IRS, and the performance gain increases as $M$ increases (e.g., by $38.82\%$ and $25.07\%$ with $M=20$ for $N=8$ and $N=32$, respectively). Moreover, the proposed algorithm also outperforms all benchmark schemes with IRS considerably, except the capacity upper bound. On the other hand, it is observed that the performance gap of our proposed algorithm from the capacity upper bound generally increases as $N$ and/or $M$ increases, which is due to the fact that in such cases, the upper bound has more DoF for the reflection coefficient design. This reveals a fundamental limitation of IRS-aided MIMO-OFDM communication due to the lack of frequency-selectivity at the IRS, which thus requires future research to overcome this issue.
\vspace{-3mm}
\section{Conclusions}\label{sec_conclusion}
\vspace{-2mm}
This paper studied the capacity maximization problem for IRS-aided point-to-point MIMO communication via joint IRS reflection coefficients and transmit covariance matrix optimization. Under frequency-flat channels, an alternating optimization algorithm was proposed to find a locally optimal solution by iteratively optimizing one optimization variable (i.e., the transmit covariance matrix or one of the reflection coefficients) with the others being fixed, for which the optimal solutions were derived in closed-form.  Moreover, alternative algorithms with lower complexity were also proposed for the asymptotically low-SNR and high-SNR regimes as well as MISO/SIMO channels. Furthermore, a MIMO-OFDM system was considered under frequency-selective channels, where a common set of reflection coefficients needs to be designed for all subcarriers. A new alternating optimization algorithm was proposed to iteratively optimize the set of transmit covariance matrices over different subcarriers or a common reflection coefficient for all subcarriers, by leveraging the convex relaxation technique. It was shown via extensive numerical results that our proposed algorithms achieve superior rate performance over various benchmark schemes with or without IRS. Moreover, it was revealed that by judiciously designing the IRS reflection coefficients, the IRS-aided MIMO channel can be significantly improved in terms of channel power, rank, or condition number, thereby leading to enhanced capacity.

\vspace{-3mm}
\bibliographystyle{IEEEtran}
\bibliography{MIMO}
\end{document}